\documentstyle[12pt]{article} \textwidth    165mm    \textheight 210mm
\begin{document}

\centerline{IBR   PREPRINT    TH-96-S033,   Sept.3,   1996 }
\centerline{PACS   04.60.+n; 03.65.-w,11.10.-z}

\vspace*{1cm}
\begin{center}
{\large \bf   CLASSICAL AND OPERATOR   ISOMINKOWSKIAN UNIFICATION OF
GENERAL AND SPECIAL RELATIVITIES FOR MATTER AND THEIR\\ ISODUALS FOR
ANTIMATTER}
\end{center}

\vspace*{0.5cm}
\begin{center}
{\large \bf Ruggero Maria Santilli} \\ Institute  for Basic Reseach \\
P.O.Box 1577, Palm Harbor, FL 34682, U.S.A. \\
E-mail ibr@gte.net Web site http://home1.gte.net.net/ibr/
\end{center}

\begin{abstract}
We recall that the Minkowskian geometry possesses basic units of space
and time which are invariant under the Poincar\'{e} symmetry. We then
show
that, by comparison, the Riemannian geometry possesses space-time
units which are not invariant under the symmetries of the
Riemannian line element, thus causing evident physical ambiguities.
We therefore introduce a novel formulation of
general  relativity in the {\it  isominkowskian geometry} which is an
axiom-preserving   lifting   of  the   conventional   Minkowskian
geometry but which nevertheless admits all possible Riemannian metrics
thanks to a (positive-definite) $4 \times 4$  generalization of the
basic unit. We construct the universal symmetry of the isominkowskian
line
elements called {\it isopoincar\'{e} symmetry},
prove that it is locally isomorphic to the conventional
Poincar\'{e} symmetry, and show that, in this way, conventional
Riemannian metrics and related field equations can be expressed with
respect to invariant generalized units of space and time.  We then show
that
the isominkowskian geometry and related isopoincar\'{e} symmetry permit:
I) A  classical geometric unification of the general and special
 relativities for matter into a formulation called
{\it isospecial relativity}  in  which the former  occurs  for
generalized units while admitting the latter as  a particular case for
conventional units; II) A novel operator
formulation of gravity for matter based on the abstract axioms
of {\it relativistic} quantum mechanics, thus showing hope for a
possible
resolution of the ambiguities in current theories of quantum gravity;
and
III) A novel classical and operator formulation of antimatter which is
an
antiautomorphic image of the preceding formulations for matter
constructed via
a map called {\it isoduality}. The  experimental validity  of  the
classical isominkowskian formulation of  gravity for matter
is derived from  the
preservation  of  the conventional   Einstein's field equations
except for inessential multiplicative terms.  The
experimental verification of  the operator  isominkowskian formulation
of  gravity for matter is derived from the
preservation of conventional quantum
mechanical laws, with  gravitational  effects  being notoriously  very
small as compared to those of other interactions of the particle word.
The
validity of the isodual formulations for antimatter is
inferred from its compatibility with available experimnental data. The
results of this paper have been made possible thanks
to the recent achievement of
sufficient maturity for mathematical content in memoir [3f],
axiomatiic consistency in memoir [3g] and generalized symmetry
principles
in memoir [7a]. Further studies, such as the formulation of an
isotopic grand unified
theory inclusive of gravitation, are presented in the fortcoming paper
[10].
\end{abstract}

\section{Introduction}
\setcounter{equation}{0}

As it is well known, the {\sl special relativity }~[1] constitutes
one of  the most majestic  scientific achievements of this century for
mathematical   beauty,     axiomatic consistency    and    unambiguous
experimental verifications.  By comparison, despite equally historical
advances through  this century,  the {\sl  general relativity}~[2]
still remains  afflicted by basic unresolved  problematic aspects.  In
this  note  we therefore     initiate  studies aimed at    a geometric
unification of the general  and  special relativity via  the  abstract
axioms of the {\sl special} rather than of the general relativity.

Our central  methodological   tools for the characterization  of  {\sl
matter} are the so--called  {\sl isotopies}~[3] which, for the case at
hand,
are characterized by the lifting of the unit of relativistic
theories $I = Diag. (1, 1, 1, 1)$ into a well behaved and
positive--definite
but otherwise arbitrary $4 \times 4$ matrix $\hat {I}(x, \dot{x},
\ddot{x}, ...)
= 1/\hat {T}$ with associated lifting of the conventional
associative product $A \times B$ among generic quantities A, B into the
isoproduct $A \hat {\times} B = A \times \hat {T} \times B$ under which
$\hat {I}$ is
the correct left and right unit of the new theory.

For consistency the entire
mathematical and physical structures of the original theories,
must be reconstructed with respect to thje above generalized unit and
product,
yielding  the so-called
{\it isonumbers, isospaces, isoalgebras, isogeometries}, etc. [3f]. It
is
easy to see that, for positive--definite generalized unit $\hat {I}$,
all
isotopic structures are locally isomorphic to the original ones and, in
this
sense, all isotopies are {\it axiom-preserving}. We should therefore
indicate from the outset that {\it the isotopies do not produce new
theories,
but merely new realizations of existing theories}, which is the main
line of study of this paper.

The isotopic structures
which are particularly significant for this paper are:
the  {\it isominkowskian  spaces}~[4a];  the  {\it
isolorentz}~[4a]  and  {\it   isopoincar\'{e}  symmetries} [4e,4f];
and the {\it  isospecial relativity}~[4], which is the axiom-preserving
formulation  of the conventional  relativity on  isominkowskian spaces
under  the isopoincar\'{e} symmetry (see~[4h]  for  a   general
presentation).

The   methods for   the  characterization of {\it
antimatter}  are the so-called  {\it isodualities }~[4b,5]
which are characterized
by the {\sl antiautomorphic} map of all quantities A for matter into
their anti-Hermitean forms $-A\dagger$, thus implying {\it
negative-definite}
units $-\hat {I}$. Again, for consistency the isodual map must be
applied to
all mathematical and physical formulations for matter, yielding {\
isodual
isonumbers, isodual isospaces, isodual isoalgebras}, etc. The isodual
structures
which are particualrly important for this paper are:
the {\it isodual  isominkowski} spaces,
the  {\it isodual isopoincar\'{e}  symmetry}, and the {\it isodual
isospecial
relativity} (see~[4h] for a general presentation).

Independent  reviews   and developments  can  be found   in
monographs~[6],
papers~[7] and literature quoted therein.
 Ref.~[6e] provides a comprehensive bibliography
up to 1984, while  a subsequent bibliographical (and technical) survey
is available in monograph~[6d].

The main lines  of the {\sl  classical}  geometric unification of  the
special and general relativities were first submitted in ref.~[4d]
as a natural  consequence of the  isopoincar\'{e} symmetry. The main
lines
of the {\sl operator} geometric unification of the special and general
relativities  were submitted for the first  time at the {\sl VII Marcel
Grossmann  Meeting on General Relativity}  held at Stanford University
in July 1994~[8a].

The above studies still lacked a rigorous form-invariant character
because they were based on the {\it conventional} differential calculus
which has resulted to be noninvariant, and thus inapplicable under
isotopies.

In this paper we present for the first time the fully form-invariant
formulations of:

I) The  classical geometric unification of the general and special
 relativities for matter into the  isospecial relativity
in  which the former  occurs  for
generalized units while admitting the latter as  a particular case for
conventional units;

II) The operator
formulation of gravity for matter based on the abstract axioms
of {\it relativistic} quantum mechanics, thus showing hope for a
possible
resolution of the ambiguities in current theories of quantum gravity;
and

III) The classical and operator isodual formulations of antimatter.

 The  experimental validity  of  the
classical isominkowskian formulation of  gravity for matter
is derived from  the
preservation  of  the conventional   Einstein's field equations
except for inessential multiplicative terms.  The
experimental validity of  the operator  isominkowskian formulation
of  gravity for matter is derived from the
preservation of conventional quantum
mechanical laws, with  gravitational  effects  being notoriously  very
small as compared to those of other interactions of the particle word.
The
validity of the isodual formulations for antimatter is
inferred from its compatibility with available experimnental data.

 The above results have been made possible by the recent achievement of
sufficient maturity for: mathematical content in memoir [3f]
including the isotopies and isodualities of differential calculus and
their
applications to algebras, geometries and analytic mechanics;
general axiomatiic consistency in the physical formulations of
both isotopic and isodual theories in memoir [3g]; and generalized
symmetry principles for isotopic and isodual theories
in memoir [7a]. Further studies, such as the formulation of a grand
unified
theory with an axiomatically consistent inclusion
of gravitation are presented in the fortcoming paper [10]. An
introductory outline
of the main mathematical and physical aspects of this paper
is available in Pages 18, 19 of Web Site [7u].

A primary motivation      for     this paper is      the    following:
\newtheorem{teorema}{Theorem}
\begin{teorema}
The basic unit of all  (nowhere degenerate, real valued and symmetric)
geometries with   non--null curvature over  conventional fields  is not
invariant under  the symmetry of their  line element in both classical
and quantum formulations.
\end{teorema}

{\bf   Proof.} Let  $E  = E(x,\delta,R)$   and  $\Re = \Re(y,g,R)$  be
n--dimensional Euclidean and  Riemannian spaces, respectively, with the
same signature  $(+,+,...,+)$,  basic unit $I  =  $diag.$(1,1,...,1)$,
metrics  $\delta  = (\delta_{ij}  = Diag.(1,  1, ...,  1)$ and $g(y) =
(g_{ij}) = g^t$, and local coordinates
$x = \{x^k\}, y  = \{y^k\}, i, j, k
= 1,2, ...n$, over the field $R = R(n,+,\times)$ of
real numbers n
with conventional sum + and multiplication $\times$.

The transformation $x \rightarrow y(x)$ for which the Euclidean metric
is mapped into the Riemannian metric,

\begin{eqnarray}
\delta_{ij}  \rightarrow g_{ij}(y) = \frac{\partial y^r}{\partial x^i}
\delta_{rs}\frac{\partial y^s}{\partial x^j},
\label{eq:one-1}\end{eqnarray}
is {\sl necessarily noncanonical } for non--null curvature.  Therefore,
the  symmetries  of the  Riemannian line elements   $y^2  = y^tgy$ are
necessarily {\sl noncanonical}.   As such,  these  symmetries do   not
generally preserve the basic unit I at the classical level because, by
definition,   noncanonical  transforms  do  not  leave   invariant the
fundamental canonical (symplectic) tensor

\begin{eqnarray}
(\omega_{\mu\nu})  =  \left( \begin{array}{cc} 0 & -I\\ I  &    0
\end{array} \right),
\label{eq:one-2}\end{eqnarray}
i.e., they are of the type in phase  space with local coordinates $b =
\{x,p\} \rightarrow b'(b) = \{x',p'\}$

\begin{eqnarray}
\omega_{\mu\nu} \rightarrow \omega'_{\mu\nu} = \frac{\partial
b^a}{\partial  b'^\mu}\omega_{\alpha\beta}\frac{\partial
b^\beta}{\partial b'^\nu} = \omega'_{\mu\nu}(a') \neq \omega_{\mu\nu}
\label{eq:one-3}\end{eqnarray}

But the canonical tensor represents the fundamental space units of the
theory, and this establishes  the inability of  classical conventional
geometries with non--null curvature to have invariant basic units.

The  symmetries of the  same line  element  in operator formulation are
then necessarily {\sl nonunitary}  for consistency (see Sect.  5), and
this proves the lack of invariance of the basic unit also for operator
theories. The same  proof evidently applies  for indefinite signatures
$(+,  +, ...  -, -....)$ (see  later  on for  the  (3+1)-- dimensional
case). {\bf q.e.d. }

\vskip 1cm
To  understand the implications of  the above theorem, recall that the
basic unit of the (3+1) -- dimensional Riemannian geometry is given by
$I = Diag. ((1, 1, 1 ), 1)$,  where the first  three components $(1,
1, 1)$ represent  the space units (say 1  cm)  in dimensionless form,
and
the  fourth  component represents  the time unit  (say  1 sec) also in
dimensionless form.

Theorem  1  establishes   that {\sl  curvature  implies   the lack  of
invariance of the fundamental  space--time  units of the theory},  thus
implying evident  problematic aspects in the  comparison of the theory
with experimental data. In fact, one of the fundamental conditions for
the applicability    of  the  measurement  theory  is   precisely  the
invariance of the basic unit.

As a result, Theorem 1 provides a new perspective of the various
problematic
aspects voiced on Einstein's gravitation during this century
(see, e.g., Ref. [5h] for an outline with referenbces) because it
indicates
that they are not necessarily due to Einstein's field equations
but rather to  their referral to  a geometry  in which the
basic units opf space and time are not invariant.
In fact, following Theorem 1,
the  same problematic aspects  can  be   proved
to   persist  for  all    possible
modifications--enlargement of Einstein field equations.

Equivalently, Theorem 1 establishes  that {\sl Riemannian spaces are a
noncanonical  deformation of  the  Minkowskian  spaces} (because   the
former are obtainable via noncanonical  transformations of the latter)
and,  as such, they suffer  of all drawbacks  of noncanonical theories
when formulated on conventional spaces over conventional fields.

In the next section we present a formulation of classical gravitation
for matyter which preserves
Einstein's gravitational field  equations unchanged and
merely reformulates them
in a  new geometry, the isominkowskian  geometry, with generalized yet
invariant  basic units under its  universal iso\-poin\-ca\-r\`{e}
symmetry. The operator and antimatter profiles are studied in subsequent
section. To render this paper selfsufficient as well as for notational
purposes, each section contains an outline of the new methods used
therein.

\section{ Classical isominkowskian  unification of the  special and
general
relativities for matter}
\setcounter{equation}{0}

As  it is well  known,  there cannot be   really new physical advances
without   new mathematics.    It  turn, there  cannot   be  really new
mathematics without new numbers.  Still in turn, the  only possibility
of  identifying   new numbers  known   to  the  author   is  via   the
generalization of their basic unit   $I = +1 \rightarrow \hat{I}$
called  {\sl lifting}  (first
proposed in~[3a,3b]), where  $\hat{I}$ is
in general a well behaved, $n \times n$ matrix
with  an  unrestricted   functional  dependence of their elements on
coordinates $x$, their  derivatives of arbitrary  order and any needed
additional quantity,

\begin{eqnarray}
I = + 1 \rightarrow \hat{I} = \hat{I}(x, \dot{x}, \ddot{x}, ...).
\label{eq:two-1}\end{eqnarray}

The  fundamental quantities of this paper  are therefore given by {\sl
new numbers   with arbitrary  units  } studied    in details  in
ref.s~[3e,3f].

A mere inspection of lifting~(\ref{eq:two-1}) reveals a  significant
broadening  of   the  conventional numbers    and,  therefore, of  the
mathematical and physical  theories built on  them.  In fact,  we have
the  following  primary classification~[3f]: 1)   Ordinary {\sl
numbers} occurring for $\hat{I}  = +1$; 2){\sl isonumbers  } occurring
for  {\sl Hermitean}  generalized  units  $\hat{I} = \hat{I}^\dagger$;
3){\sl genonumbers} occurring for {\sl  nonhermitean } generalized unit
$\hat{I}\neq \hat{I}^\dagger$; and 4){\sl hypernumbers} occurring
for generalized units given by an {\sl  ordered set of Hermitean
quantities}.   The latter three classes   then admit
subclasses depending on whether the (real part of the)generalized unit
is   positive--definite,   singular,  etc.   The  numbers,  isonumbers,
genonumbers   and hypernumbers are used   for  the description of {\sl
matter} in  condition  of progressively increasing   complexity (e.g.,
reversible,  nonreversible,  etc.), with   the  most general  possible
hypernumbers  being studied for the description  of matter in its most
complex conditions (e.g., the DNA code).

Moreover,   the   conventional   numbers    and each    of  the  above
generalizations   admit  antiautomorphic   images  $\hat{I}\rightarrow
\hat{I}^d = -\hat{I}^\dagger$ called {\sl isoduality}~[4b,3e,3f]
which  are used  for   the description of    {\sl antimatter} also  in
physical conditions of progressively increasing complexity [5].

It is evident  that in a  field of such a diversity  we are forced for
brevity  to restrict our studies    to  the first  lifting, those   of
isotopic type  and its  isodual.  In  this section we  shall therefore
study  a classical representation  of matter  based on the isonumbers,
the corresponding isodual representation of  antimatter is studied  in
the    next section.  Corresponding   operator  images  are studied in
subsequent   sections. The genotopic  and hyperstructual extensions  of
isotopic   formulation of this  paper  are  contemplated for study  in
subsequent works.

Let  $F = F(a,+,*)$ be a  field of numbers (i.e.,  real numbers $a = n
\in R$, complex numbers $a  = c \in C$  or quaternions $a  = q \in Q$)
with   conventional sum $a +  b$  and  product  $a\times b   = ab$ and
corresponding additive unit 0 and multiplicative unit I.

The {\sl isofields}  are rings $\hat{F}  = \hat{F}(\hat{a},+,\hat{x})$
with elements $\hat{a}   =   a\times\hat{I},  a\in F$   called    {\sl
isonumbers},where    $\hat{I}$ a  positive--definite quantity   (e.g., a
matrix) generally {\sl outside} the original set $F$ equipped with:

\begin{itemize}
\item[i)] the {\sl isosum} $\hat{a}+\hat{b} = (a+b)\times\hat{I}$ with
conventional additive unit $\hat{0} \equiv 0, \; \hat{a} + \hat{0}
\equiv
\hat{0} + \hat{a}   \equiv\hat{a}\: \in F  $  (the preservation  of the
additive unit 0 is indicated by preserving the symbol $+$ unchanged in
$\hat{F}(\hat{a},+,\hat{x})$); and

\item[ii)]the {\sl isoproduct }
\begin{eqnarray}
\hat{a} \hat {\times} \hat{b} =   \hat{a}\times  \hat{T} \times \hat{b}
=   ( a
\times b ) \times \hat{I} \in \hat{F},
\label{eq:two-2}\end{eqnarray}
under which the quantity $ \hat{I}  = \hat{T}^{-1}$ is the correct new
left and right unit of $\hat{F}$

\begin{eqnarray}
\hat{I}  \hat{\times} \hat{a}   \equiv   \hat{a} \hat{\times}  \hat{I}
\equiv \hat{A}, \forall \hat{a} \in \hat{F}, \hat{I} = \hat{T}^{-1},
\label{eq:two-3}\end{eqnarray}
\end{itemize}
(the change  of the   multiplicative  unit is indicated with   the new
symbol  $\hat{\times}$ in $\hat{F}(\hat{a},+,\hat{\times})$).  When the
above conditions are verified,    the  $\hat{I}$ is called   the  {\sl
isounit} and $\hat{T}$ is called the {\sl isotopic} element.

The fundamental mechanism  of   the isotopies responsible   for  their
axiom--preserving character is that of lifting the multiplicative unit
$ I \rightarrow \hat{I}$ while  jointly the product  is lifted by  the
{\sl    inverse}    amount,   $\times  \rightarrow   \hat{\times}    =
\times\hat{T}\times,   \hat{I}   = \hat{T}^{-1}$.   This  implies  the
following

{\bf Lemma 1~[3e]:}  {\sl Isofields satisfy all  the axioms of a
field.}

Despite the  local isomorphism $\hat{F}   \approx  F$, the lifting  $F
\rightarrow  \hat{F}$   is   not  trivial   because  it    requires  a
corresponding lifting of   all  operations of $F$. For   instance, the
conventional square of a number  $n^2 = n\times  n$ has no meaning for
$\hat{F}$ and must be lifted into the {\sl isosquare} $\hat{n}^2 =
\hat{n}
\hat{\times} \hat{n}$. Along similar lines we have the {\sl isopowers}
$\hat{n}^{\hat{m}}        =     \hat{n}\hat{\times}\hat{n}\hat{\times}
.. \hat{\times} \hat{n}  $   (m  times);  the {\sl  isosquare   root}
$\hat{n}^{\hat{\frac{1}{2}}}                                         =
n^{\frac{1}{2}}\hat{\times}\hat{I}^{\frac{1}{2}}$;       the      {\sl
isoquotient } $\hat{n}\hat{/}m = (\hat{n}/\hat{m})\times \hat{I}$; the
{\sl isonorm} $\hat{|}\hat{a}\hat{|} = |  a | \times \hat{I}$ ,  where
$| a |$ is the conventional norm; etc. (see~[3e] for brevity for
all details).      The nontriviality of   the  lifting  $F \rightarrow
\hat{F}$ is then illustrated by the fact  that numbers which are not
prime
for   $I  = +1$   may become prime  for other units~[3e].

The axiomatic  consistency   of the  emerging new   structure  is
established by the local isomorphism between the conventional field and
its
isotopic image. In particular, it should be notes that
{\sl  the isounit preserves all axiomatic
properties of the original unit}, e.g.,

\begin{eqnarray}
\hat{I}^{\hat{m}}  =       \hat{I} \hat{\times}   \hat{I} \hat{\times}
.. \hat{\times}  \hat{I}  \equiv \hat{I}, \,
\hat{I}^{\hat{\frac{1}{2}}}
\equiv   \hat{I};     \,   \hat{I}/\hat{I}   \equiv      \hat{I},   \,
\hat{|}\hat{I}\hat{|} \equiv \hat{I}, \, {\rm etc}.
\label{eq:two-4}\end{eqnarray}

The assumption of the isonumbers as the fundamental numbers requires a
simple, yet  unique and    unambiguous reconstruction of
contemporary mathematical  and physical theories into  forms admitting
of $\hat{I}$, rather than $I$, as the correct left and right unit.

Recall that the conventional metric spaces of contemporary physics are
based on conventional  fields of number. the  first and most important
implication of the  lifting  $I \rightarrow \hat{I}$ is  therefore the
necessity  to  construct,   for   evident  reason   of    consistency,
corresponding liftings of metric spaces.

Let $M  = M(x, \eta, R)  $ be a conventional Minkowski space~[1]
with coordinates $x = \{x^\mu\} = \{r, c_{o}t\}, \mu =  1, 2, 3, 4,$
where
$c_{o}$ is the   speed of light  in  vacuum,  with  basic unit  $  I =
Diag. (+1, +1, +1, +1) $  and metric $\eta =  Diag. (+1, +1, +1, -1) $
over a field   $R = R(n,+x)$ of real   numbers $n$  equipped with  the
conventional   sum $+$ and product   $\times$,  additive unit $0$  and
multiplicative   unit  $I$.

 The   lifting  $R(m,+,\times) \rightarrow
\hat{R}(\hat{n},+,\hat{\times})$  then     requires  the corresponding
lifting of  the Minkowski  into  the {\sl  isominkowski  spaces} $M  =
M(x,\eta,R) \rightarrow \hat{M} = \hat{M}(\hat{x},\hat{\eta},\hat{R})$
first submitted  in~[4a]   which  are  characterized by    the   {\sl
isocoordinates} $\hat{x}  =  x\times\hat{I}$ on   $\hat{R}$, the {\sl
isometric}  $\hat{N}_{\mu\nu}   =    \hat
{\eta}_{\mu\nu}\times\hat{I}      =
\hat{T}_{\mu}    ^{\alpha}(x,              \dot{x},          \ddot{x},
..  )\eta_{\alpha\nu}\times\hat{I}$ where $\hat{I} = \hat{T}^{-1}$ is
now  in general a $4\times 4$  positive--definite matrix, and the {\sl
isoseparation} is hereon expressed for diagonal isounits

\begin{eqnarray}
\hat{M}(\hat{x}, \hat{\eta}, \hat{R}) : \hat{x} = x \times \hat{I},
\hat {eta} = \hat{T}(x, \dot{x}, \ddot{x}, ...) \times \eta, \, \hat{I}
= \hat{T}^{-1},
\label{eq:two-5a}\end{eqnarray}
\begin{eqnarray}
(\hat{x}   -   \hat{y})^{\hat{2}}    & = &  (\hat{x}^\mu   -\hat{y}^\mu)
\hat{\times} \hat{N}_{\mu\nu} \hat{\times}(\hat{x}^\nu  - \hat{y}^\nu)
= [(x^\mu - y^\mu) \times \hat{\eta}_{\mu\nu}  \times (x^\nu - y^\nu)]
\times  \hat{I}   = \nonumber \\
&=& [(x^1    - y^1) \times T_{1  1}(x,  ...) \times (x^1-y^1)  +
(x^2-y^2) \times T_{2   2}(x,  ...) \times(x^2-y^2)   +   \nonumber \\
& & +  (x^3-y^3) \times T_{33}(x,
..) \times (x^3-y^3) - (x^4-y^4) \times T_{4 4}(x, ...) \times
(x^4-y^4)] \times \hat{I},
\label{eq:two-5b}\end{eqnarray}
\begin{eqnarray}
\hat{T} =  Diag.(T_{1  1}, \hat{T}_{22}, \hat{T}_{33},  \hat{T}_{44}),
\hat{T}_{\mu\mu}> 0, \;\; x, y \in M, \hat{I} \not\in M.
\label{eq:two-5c}\end{eqnarray}

  Note that all  scalars of  M must  become  {\sl isoscalar} to   have
meaning   for  $\hat{M}$, i.e., they must    have the structure of the
isonumbers $\hat{n} =   n\times\hat{I}$.  This condition  requires the
re--definition       $x    \rightarrow   \hat{x}     = x\times\hat{I},
\hat{\eta}_{\mu\nu}          \rightarrow      \hat{N}_{\mu\nu}       =
\hat{\eta}_{\mu\nu}\times\hat{I}$, etc.

Note however the practical redundancy of using isocoordinates $\hat{x}
= x\times\hat{I}$.   In   fact,  we can   write  $\hat{x}^{\hat{2}}
 = (x^\mu\times\hat {\eta}_\nu\times x^\nu)\times\hat{I}  =
x^{\hat{2}}$.  For  simplicity  we shall hereon   use the conventional
coordinates. Note also the redundancy of using the full isoscalar form
$\hat{N}$ of the isometric because  the reduced form $\hat{\eta}$ with
ordinary  elements  $\hat{\eta}_{\mu\nu}$ in   $R$ is  sufficient. The
understanding is that  the full isotopic  formulations  are needed for
mathematical consistency.

We   shall hereon  assume    the convention,  rather  familiar in  the
literature  of the isotopies, that   all  quantities with a "hat"  are
computed in isospaces over isofields, and the corresponding quantities
without a "hat" are computed  on conventional spaces over conventional
fields.

Note the necessary condition  that {\sl isospaces and isofields  have
the same isounit}.   This  condition is  absent   in  the conventional
Minkowski space where the unit of  the space is  the unit {\sl matrix}
$I = diag.  (1, 1, 1, 1)$  while that of the  underlying field is  the
{\sl number}  $I =  +1$. Nevertheless,   the latter  can  be trivially
reformulated with the common unit matrix $I$, by achieving in this way
the form admitted as a particular case by the covering isospaces
\begin{eqnarray}
M(x, \eta, R)  : \, x^2 = (x^\mu\times\eta_{\mu\nu}\times x^\nu)\times
I \in R
\label{eq:two-6}\end{eqnarray}
Also, one should keep in mind for future needs the following

\begin{eqnarray}
{\rm Basic-Isoinvariant = (length)^2 \times (unit)^2}.
\label{eq:two-7}\end{eqnarray}

A fundamental  property of the infinite  family  of generalized spaces
(8) is that the lifting of the basic unit $I\rightarrow
\hat{I}$ while the metric is lifted of the {\sl inverse} amount, $\eta
\rightarrow  \hat{\eta} =  \hat{T}\times\eta, \hat{I} = \hat{T}^{-1}$,
implies   the preservation  of all   original  axioms, and  we have the
following:

\vskip  1cm  {\bf Lemma 2~[4]:}   {\sl The isominkowski  spaces
$\hat{M}(\hat{x},  \hat{\eta},   \hat{R})$      over   the   isofields
$\hat{R}(\hat{n},+,\hat{\times})$   with  a common  positive--definite
isounit $\hat{I}$ preserve all  original axioms of the Minkowski space
$M(x,\eta,R)$ over the reals $R(n,+,\times)$.}

\vskip 1cm

The  nontriviality  of the  lifting   is   that  of gaining   an  {\sl
unrestricted functional   dependence  of   the metric    $\hat {eta}  =
\hat{\eta}(x,  \dot{x}, \ddot{x})$  under the conventional Minkowskian
axioms.}

The local isomorphism $\hat{M} \approx M$ holds to such an extent that
   the   {\sl isominkowski and    Minkowski   spaces coincide at   the
   abstract,  realization--free level by  conception and construction.}
   Thus, {\sl the isominkowskian spaces are not new spaces, but merely
   "new realizations" of the original abstract Minkowskian axioms}.  In
   particular, the maximal causal  speeds of both spaces coincides and
   it is given by $c_0$ as we shall see.

Lemma   2 illustrates again  the  "axiom--preserving"  character of the
isotopies  indicated  in Sect.1,  this  time  at the  level  of metric
spaces.  It  should  be    stressed  that the "isotopies"  are    {\sl
inequivalent} to the various forms   of "deformations" of the  current
literature  for   several  reasons,   such     as:  the former     are
axiom--preserving while the latter are not; the former are defined over
generalized fields   while the latter are  not;   etc.  {\sl  To avoid
confusion, readers are discouraged  from using the term "deformations"
(of given  structure into a nonisomorphic form)  when referring to the
"isotopies" ( of the same  structures into axiom--preserving isomorphic
forms).}

The   {\sl   isominkowskian geometry} was    first   proposed in
ref.~[4a]
(see  ref.s~[4g,4h] for  comprehensive  studies  and
ref.s~[6,7] for independent  works).  These studies were however
incomplete because based  on  the  conventional  differential calculus
which  has  resulted   to be   inapplicable    under isotopies.    The
foundations of   the   isominkowskian geometry   formulated   via  the
isodifferential   calculus   of   ref.~[3f]  are introduced  here
apparently  for the first time,   with more detailed studies presented
elsewhere~[9].

Stated in a nutshell, the isominkowskian geometry is a symbiotic union
of the Minkowskian and  Riemannian   geometries along the   following
main
properties:

\vskip 1cm    {\bf  I) Isoflatness.} It is easy to see that the
isominkowskian geometry  is {\sl flat} in  isospace over
isofields, a property called
{\sl  isoflatness}~[4g]. This is due to the fact that curvature, which
is
represented by the factor $\hat {T}$ in the isometric $\hat {eta} =
\hat {T} \times {eta}$, is referred to its own inverse as unit.
In fact, the  new geometry permits the
definition in  isospace of   straight line and   intersecting  angles,
although  in a predictable generalized form,  which is  not possible in
the Riemannian geometry, thus confirming the preservation of the {\sl
Minkowskian} axioms.

In particular,  isoflatness   allows the  reconstruction in   isospace
$\hat{M}  (\hat {x},\hat{\eta},\hat{R})$ of the trigonometric,
hyperbolic and
other functions {\sl  for   a  metric with an   arbitrary   functional
dependence}, which we cannot possibly  review here (see~[4g] for
brevity).

More generally,  the isominkowskian geometry  is based on the new {\sl
isofunctional  analysis}   in  which  the   isospace $\hat{M}(\hat{x},
\hat{\eta},\hat{R})$  is turned into   an {\sl isomanifold}  thanks to
{\sl Kadeisvili's isocontinuity}~[7p]  and {\sl  Tsagas--Sourlas
isotopology}~[7q].  Within such  a setting, the isogeometry must
be   solely elaborated  with  all   special isofunctions,
isotransforms  and
isodistributions, etc.  (see~[4g,4h] for details).

\vskip 1cm   {\bf  II)  Pseudocurvature.} In  view  of   the arbitrary
functional dependence  of  the  isometric $\hat{\eta} =  \hat{\eta}(x,
\dot{x}, \ddot{x})$,  the isominkowskian geometry can  also be
considered as being {\sl  curved}, but only
when  projected in   the original  space  M over  the
original  field R, a property  is called {\it pseudocurvature}
which  is of  {\sl Riemannian} (rather than Minkowskian) character.

This illustrates the symbiotic capacity of the isominkowskian geometry
of  unifying   the   main characteristics of    both, the  Minkowskian
and
Riemannian geometries.  In  turn, such character  is  evidently at the
foundation of  the proposed isominkowskian  unification of the special
and general relativities.

To outline the pseudocurved character, consider the isominkowskian {\sl
manifolds} $\hat{M}(\hat{x},\hat{\eta},\hat{R})$     equipped      with
Kadeisvil's  isocontinuity and  Tsagas-Sourlas isotopology.  The  {\sl
isodifferential calculus } on  $\hat{M}(\hat{x}, \hat{\eta}, \hat{R})$
  can be   defined via the  following
notions of {\sl isodifferentials} and {\sl isoderivatives}~[3f]
\begin{eqnarray}
\hat{d}x^{\mu}    =       \hat{I}^{\mu}_{\nu}\times          dx^{\nu},
\hat{\partial}_{\mu} =    \hat{\partial}   / \hat{\partial}x^{\mu}   =
\hat{T}_{\mu}^{\nu}\times                 \partial_\nu               =
\hat{T}_\mu^\mu\times\partial/\partial x^\nu,
\label{eq:two-8a}\end{eqnarray}

\begin{eqnarray}
\hat{\partial}x^\mu  /    \hat{\partial}x^\nu  =   \delta^\mu_\nu,  \,
\hat{\partial}x_\mu / \hat{\partial}x^\nu = \hat{\eta}_{\mu\alpha}  \,
\hat{\partial}x^\alpha / \hat{\partial}x^\nu = \hat{\eta}_{\mu\nu},
\label{eq:two-8b}\end{eqnarray}
and other axiom-preserving properties here omitted for brevity, where
we have ignored for notational simplicity the isoquotient and related
factorization of the isounit. In this way,
$[\hat{\partial}x_\mu \hat{/} \hat{\partial}x^\nu] \hat{\times} F =
[\hat{\partial}x_\mu / \hat{\partial}x^\nu] \times F$.

Since the isometric $\hat{\eta}$ has an explicit  dependence on x, the
  isominkowskian geometry does indeed allow   the introduction of  the
  following {\sl isoconnection}, called {\sl isochristoffel's symbols}

\begin{eqnarray}
\hat{\Gamma}_{\alpha\beta\gamma}  =  \frac{1}{2}(\hat{\partial}_\alpha
\hat{\eta}_{\beta\gamma}                                             +
\hat{\partial}_\gamma\hat{\eta}_{\alpha\beta}                         -
\hat{\partial}_\beta\hat{\eta}_{\alpha\gamma})      =    \hat{\Gamma}_
{\gamma\beta\alpha},
\label{eq:two-9a}\end{eqnarray}

\begin{eqnarray}
\hat{\Gamma}_{\alpha }^{\beta }{}_{\gamma } =   \hat{\eta}^{\beta\rho}
\times   \hat{\Gamma}_{\alpha\rho\gamma}     =   \hat{\Gamma}_{\gamma}
^{\beta}{}_{\alpha},         \,       \hat{\eta}^{\beta\rho}         =
[|\hat {\eta}_{\mu\nu}|^{-1}]^{\beta\rho}.
\label{eq:two-9b}\end{eqnarray}

The  {\sl isocovariant differential  } of a  vector  field can then be
defined by

\begin{eqnarray}
\hat{D}    \hat{X}^\beta    =       \hat{d}     \hat{X}^\beta        +
\hat{\Gamma}_\alpha^\beta{}_\gamma
\hat {\times} \hat{X}^\alpha \hat {\times} \hat{d}\hat{x}^y,
\label{eq:two-10}\end{eqnarray}
where isoproduct can be reduced to ordinary ones because of
the cancellation of $\hat{I}$ and $\hat {T}$,
with corresponding {\sl isocovariant derivative}

\begin{eqnarray}
\hat{X}^\beta_{\uparrow\mu}  =      \hat{\partial}_\mu\hat{X}^beta   +
\hat{\Gamma}_ \alpha^\beta{}_\mu \hat {\times} \hat{X}^\alpha,
\label{eq:two-11}\end{eqnarray}

The   isotopy  of  the  proof  of  the   conventional Riemannian
case~[11],
pp.~80--81, yields the following:

\vskip  1cm  {\bf Lemma  3 (Isoricci Lemma):}   {\sl Under the assumed
conditions,   the  isocovariant derivatives  of  all  isometrics of the
isominkowskian spaces are identically null,}

\begin{eqnarray}
\hat{\eta}_{\alpha\beta\uparrow\gamma} \equiv 0,   \:    \alpha,
\beta, \gamma = 1, 2, 3, 4.
\label{eq:two-13}\end{eqnarray}

The novelty is  illustrated by the   fact that {\sl the  Christoffel's
symbols,  the covariant derivative and  the Ricci Lemma persist under:
1) an arbitrary  dependence of the  metric $\hat{\eta} = \hat{\eta}(x,
\dot{x}, \ddot{x},  ...)$, rather than the  current restriction to the
Riemann dependence only, $g=g(x)$; 2) under the Minkowskian,
rather than Riemannian axioms; and 3) with null curvature in isospace
over
isofields. }

It should   be   noted that  the above    properties  were  studied
in~[3f] for the {\sl   isoriemannian} geometry, and the above  {\sl
isominkowskian } reformulation is submitted here for the first time.

We   now  introduce  on    $\hat{M}(\hat{x},\hat{\eta},\hat{R})$:  the
following   {\sl isocurvature  tensor,   isoricci  tensor,}  and  {\sl
isocurvature isoscalar}

\begin{eqnarray}
\hat{R}_\alpha^\beta{}_{\gamma\delta}                                =
\hat{\partial}_\delta\hat{\Gamma}_      \alpha^\beta{}_\gamma        -
\hat{\partial}_\gamma\hat{\Gamma}_\alpha^\beta{}_\delta              +
\hat{\Gamma}_\rho^\beta{}_\delta                               \hat
{\times}
\hat{\Gamma}_\alpha^\rho{}_\gamma  -  \hat{\Gamma}_\rho^\beta{}_\gamma
\hat {\times} \hat{\Gamma}_\alpha^\rho{}_\delta ,
\label{eq:two-13a}\end{eqnarray}
\begin{eqnarray}
\hat{R}_{\mu\nu} = \hat{R}_\mu^\beta{}_{\nu\beta} ,
\label{eq:two-13b}\end{eqnarray}
\begin{eqnarray}
\hat{R} = \hat{N}^{\alpha\beta} \hat {\times} \hat{R}_{\alpha\beta} .
\label{eq:two-13c}\end{eqnarray}

Einstein's field equations on isominkowskian spaces can then be written

\begin{eqnarray}
\hat{G}_{\mu\nu}  = \hat{R}_{\mu\nu} -
\hat {\frac{1}{2}} \hat {\times} \hat{N}_{\mu\nu}
\hat {\times} \hat{R} = \hat{k} \hat{\times} \hat{\tau}_{\mu\nu} ,
\label{eq:two-14}\end{eqnarray}
where   $\hat{\tau}_{\mu\nu}$     is  the    source     tensor   on  $
\hat{M}(\hat{x},\hat{\eta},\hat{R}), \hat{k}  = k\times \hat{I}$   and
$k$ the usual constant.

Despite apparent differences, it should be  indicated that
Eq.s~(\ref{eq:two-14})
{\sl  coincide numerically with  Einstein's  equations}.  In fact, the
isoderivative   $\hat{\partial}_\mu  =       \hat{T}_\mu       ^\alpha
\times\partial_\alpha$   deviates  from   the  conventional derivative
$\partial_\mu$ by the isotopic factor $\hat{T}$ (here assumed as being
diagonal).    But its numerical value  must  be referred to $\hat{I} =
\hat{T}^{-1}$, rather than $I$, thus  preserving the original value of
$\partial_\mu$.

Similarly, the    isochristoffel's symbols~(\ref{eq:two-9a}) deviate
from  the
conventional symbols by the same factor $\hat{T}$ (because $\hat{\eta}
\equiv    g$). But these   symbols  must be   referred  to the isounit
$\hat{I}$, thus  preserving conventional values.  A  similar situation
occurs    for the isocurvature   tensor~(\ref{eq:two-13a})  because the
factor
$\hat{T}$  from    the  covariant isoderivative    $\hat{\partial}$ is
compensated    by the   factor   $\hat{I}$     originating from   the
contravariant  index $\beta$,  with similar  results  holding for  the
remaining quantities.  Possible residual terms are inessential because
common factors to both sides of Eq.s~(\ref{eq:two-14}).

A  more detailed  study  of  Eq.s~(\ref{eq:two-4})  and  related
isominkowskian
geometry is  presented in  ref.~[9], including  the use  of the
forgotten {\sl Freud identity} of  the Riemannian geometry in isotopic
form.

\vskip 1cm  {\bf III) Isosymmetries.}  A  primary reason for introducing
the isominkowskian spaces and related geometry is that they permit the
construction of the universal symmetry
of the line element~(\ref{eq:two-5b}) under an {\sl
unrestricted} functional  dependence  of  the isometric $\hat{\eta}(x,
\dot{x},  \ddot{x},    ...)$~[4], while such  a   possibility is
precluded in the Riemannian geometry.

Under  the   above isotopic reformulations,  the   symmetries  of the
isoinvariant~(\ref{eq:two-5b}) can be explicitly  computed
and  are given by: the
{\sl  isorotational symmetry} $\hat{O(3)}$   [4b]for the  space
component of
isoinvariant~(\ref{eq:two-5b});  the  {\sl isolorentz  symmetry}$
\hat{L}$(3.1)~[4a];
the {\sl isotopic $S\hat{U}(2)$--spin symmetry}~[4c];
the     {\sl    isopoincar\'{e}   symmetry}    $\hat{P}$(3.1)$       =
\hat{L}$(3.1)$\times   T$(3.1)~[4d];  and the  {\sl isospinorial
isopoincar\'{e}  symmetry} $\Re$(3.1)$ =   S\hat{L}(2.\hat{C})\times
\hat{T}$(3.1)~[4e].

These {\sl isosymmetries}  are constructed via  the {\sl isotopies  of
Lie's  theory } including the  isotopies  of enveloping algebras,  Lie
algebras,  Lie   groups  transformation  and   representation theories,
etc.,
originally proposed  by the author  in memoir~[3a], developed in
various  works~[3,4,5], studied  by a  number of independent
researchers~[6,7] and  today called  {\sl  Lie--Santilli
isotheory}~[6,7].
The     isosymmetries     $\hat{O}(3),   S\hat{U}(2),
\hat{L}(3.1),  \hat{P}(3.1)$  and $\hat{\cal  P}(3.1)$ are essentially
the conventional symmetries $O(3),   SU(2), L(3.1), P(3.1)$  and ${\cal
P} (3.1)$    reconstructed   for  {\sl   arbitrary}   generalized  units
$\hat{I}(x, \dot{x}, \ddot{x},  ...)$ of the class admitted  ($4\times
4$--  dimensional, real--valued,   symmetric    and  positive--definite
matrices with the same dimension of the representation considered).

Since $\hat{I}$ is  positive--definite  by assumption, {\sl  the  above
isosymmetries are isomorphic  to the original symmetries by conception
and constructions}~[4].   As such,they are not "new symmetries",
and merely constitute the  most general known nonlinear,  nonlocal and
noncanonical {\sl realizations} of the conventional symmetries.

For classical realizations of the above isosymmerties we have to refer
the interested reader to monographs~[4g,4h]. An outline of their
operator realization is presented in Sect. 7.

\vskip 1cm
{\bf 4) Isospecial relativity.} The preceding  formalism is reduced to
primitive laws under the name of {\sl isospecial relativity}~[4]
which  is  the isotopic image    of the special  relativity~[1]
realized on isominkowskian spaces $\hat{M}(\hat{x},\hat{\eta},\hat{R})$.
  As such, we do not have a "new"
relativity, but merely the most general known {\sl realization} of the
axioms of the conventional relativity.  In fact, {\sl the special  and
isospecial relativity coincides  at the  abstract level by  conception
and construction,  to such an extent to  have the same light cone with
the same maximal causal speed $c_0$ } (see below).

\vskip 1cm
{\bf 5)  Isoanalityc mechanics.}  Finally,  the  preceding  formalism
is
complemented with  step--by--step--isotopies  of conventional
Lagrangian
and Hamiltonian  mechanics called  {\sl isoanalityc  mechanics}, which
begins with a  basic action in isospace and  includes the isotopies of
equations by Lagrange's,  Hamilton's, Hamilton--Jacobi, etc.
(see~[3f]  for
brevity).

\vskip 1cm
Despite all   these similarities   between  conventional  and  isotopic
structures, one should keep  in  mind that the conventional  Minkowski
geometry, the Poincar\'{e}  symmetry   and  the special  relativity
are
strictly linear, local--differential  and Hamiltonian. On the contrary,
their isotopic images are  generally  {\sl nonlinear} (in any  desired
variable), {\sl nonlocal--integral }  (i.e., they admit integral  terms
under  the  Tsagas--Sourlas integro--differential  topology~[5q],
provided  that they are all embedded  in  the generalized unit
$\hat{I}$),
and {\sl nonlagrangian} (in
the  sense  of admitting terms     simply beyond any  representational
capability of a (first-order) lagrangian (see later on for details).

Also, the isotopic structures have been proved by Aringazin~[7r]
to be {\sl  directly universal}, that is,  applying for all infinitely
possible, well  behaved, signature  preserving generalizations  of the
Minkowski metric (universality),  directly   in  the x--frame of    the
observer  (direct  universality).    As   a result,   any conventional
deformation  of  M   can be  {\sl   identically} reformulated  via the
isotopic formalism, and  the isotopic, axiom--preserving representation
of  deformed Minkowski spaces holds {\sl  even when not desired}, thus
implying  the existence of   various  applications  and   experimental
verifications~[7r].

The  isominkowskian  geometry,  the  isopoincar\'{e} symmetry  and
the  related
isospecial  relativity  were  originally  introduced~[4] for a
direct  representation of  {\sl   interior dynamical problems,   e.g.,
electromagnetic waves propagating  within inhomogeneous  and isotropic
physical media with a locally varying speed } $c = c_{0}/n$ as
occurring, e. g.,
in our atmosphere.  In   fact, the  first physical  application  of the
isotopic line element~(\ref{eq:two-5b})  is that of
directly representing in  its
fourth component  the local speed of light  $c = c_0/n$ with $ \hat{T}^
{\frac{1}{2}} = n_4 = n$, while the
remaining components $\hat{T}_{kk}^{\frac{1}{2}} = n_k$
     represent the anisotropy of   the  medium
considered,  and  its   inhomogeneity is  represented,  e.  g.,  via a
dependence of the quantities $n_\mu$ in the local density.

It   is evident that  the conventional   formulation  of the special
relativity in Minkowski space M  is inapplicable to (and not "violated"
by) locally varying  speeds $c = c(x, ...)$. The  use of the
isominkowskian
space permits the regaining  of the validity  of the special relativity
because in   the latter case   the speed of light  is  deformed by the
amount $c_0 \rightarrow c = c_0/n_4$
 while the related unit is deformed  by the inverse amount,
$I_{44}^{\frac{1}{2}} = 1 \rightarrow \hat{I}_{44}^{\frac{1}{2}} = n_4$,
 thus implying the constant value $c_0$ in isospace.

In  this way the special relativity   is rendered "directly universal"
under the isotopies,   that is,  it  is  rendered applicable for   all
possible local speeds of light.  Equivalently, we can   say
that the speed of light $c_0$ is a  "universal constant" only in
isospace
$\hat{M}$, while its projection in our space--time $M$ acquires the
local value
$c$.

Recall the known problematic  aspects of the  conventional formulation
of the special  relativity when applied  to physical  media, e.g., for
the  propagation  of electromagnetic  waves   in water
 where $c = c_0/n_{4} < c_0$
  and electrons can propagate with speeds {\sl greater} than the local
speed
of   light  (Cherencov  light),  thus   implying   evident  problems of
causality.  If one assumes as the  maximal causal speed  {\sl in water}
to be the  speed of light  {\sl in  vacuum}, the  principle of
causality  is
indeed  salvaged, but there  is  the loss of  the  relativistic law of
addition of speeds  because the sum of two  speeds of lights in water
does not
yield  the local speed of  light or that in   vacuum, $v_{tot} = 2c/(1 +
c/c_0) \not = c$ and $c_0, c = c_0/n$. All the above
problematic aspects  are  resolved  by the  isospecial   relativity in
isospace with a number of additional preliminary
applications and verifications (see~[4--7]for brevity).

The  isospecial   relativity was  also   introduced~[4] for  an
invariant description of  extended--deformable particles moving  within
physical media under unrestricted external  forces. This objective  is
achieved via the realization of the  isounit in the diagonal form (for
spheroidal ellipsoidical shapes)

\begin{equation}
\hat{I} = Diag.(n_1^2, n_2^2, n_3^3, n_4^2)\times\hat{\Gamma}
(x, \dot{x}, \ddot{x}, ...),
\label{eq:two-15}\end{equation}
where: a) the quantities $n_1^2, n_2^2, n_3^2, n_4^2$
 provides a geometrical representation
of the extended, nonspherical and
deformable  shape under the  volume preserving condition
$ n_1^2\times n_2^2\times n_3^2\times n_4^2 = 1$;
the quantity $n_4^2$ provides a geometric representation of the density
of
the medium in which  motion occurs, such as the  (square of the) index
of refraction;   and   the   quantity  $\hat{\Gamma}(x, \dot{x},
\ddot{x}, ...)$  represents     nonlinear,
nonlocal integral   and  nonhamiltonian interactions;    all this  in a
manifestly form--invariant way   because the isounit $\hat{I}$  is  the
basic
invariant of the isopoincar\'{e} symmetry~[4].  These features have
permitted additional applications and verifications with effects due to
deformations of shape~[4--7].

The application  of the isospecial  relativity submitted in this paper
is basically  different than the above  ones.  In fact, in  this paper
deals  with  the {\sl  classical  isotopic formulation  of the general
relativity for matter via the isominkowskian geometry}.
It should be  stress up--front
that it would be  unreasonable to expect   in this first  introductory
paper a    comprehensive treatment at   all possible  epistemological,
geometric, operator, quantum field field theoretical, experimental and
other  aspects.  The rather  limited objective   of  this study  is to
identify  the essential   axiomatic   foundations of   the  isospecial
relativity in its application to gravity,
point out its plausibility for resolving at least some of rather
old problematic aspects  of the conventional formulation of gravity, and
indicate
its experimental validity.

In summary, the representation of gravity via the isospecial relativity
is based
on the restriction of the isominkowskian metric to represent identically
any
given Riemannian metric,and in the isotopic factorization of the
{\it conventional} Minkowski metric,

 relativity is  characterized by the isominkowskian geometry
whose  isometric  $\hat{\eta}$
is   assumed  to coincide   with the  Riemannian
one, $\hat{\eta} \equiv g(x)$, and we shall write

\begin{equation}
\hat{M}(\hat{x},\hat{\eta},\hat{R}): \, \hat{\eta} \equiv g(x) =
\hat{T}_{gr}(x) \times \eta, \, \hat{I}_{gr}(x) = [\hat{T}_{gr}(x)]^{-1}
> 0.
\label{eq:two-16}\end{equation}
Note that, since the Riemannian geometry is locally Minkowskian, the
$4\times 4$ matrix $\hat{T}_{gr}(x)$ in the isominkowskian factorization

\begin{equation}
g(x) = \hat{t}_{gr}(x) \times \eta,
\label{eq:two-17}\end{equation}
is always {\sl positive--definite}. In this  case $\hat{T}_{gr}$
 and $\hat{I}_{gr}$ are called
the {\sl gravitational isotopic element and isounit}, respectively. As
an   illustration,   the   gravitational  isotopic   element   in  the
isominkowskian decomposition of the Schwarzschild's metric is given by

\begin{equation}
\hat{T}_{gr} = Diag.((1 - M/r)^{-1}, (1-M/r)^{-1}, (1-M/r)^{-1},
(1-M/r)^{-1}),
\label{eq:two-18}\end{equation}
and recovers the   conventional value  $I = Diag.(1, 1, 1, 1)$ for
$r \rightarrow \infty$.   The
isominkowskian formulation of any other model is then straightforward.

The primary motivations supporting  the isominkowskian
formulation of gravity are the following:

A)  The formulation possesses  an invariant generalized basic unit
$\hat{I}_{gr}$,
thus resolving the problematic aspects caused by Theorem 1 (Sect. 1).

B) The  formulation permits, apparently  for the  first time,
{\sl the geometric
unification  of  the special and    general relativities} into  one
single   theory,  {\sl the      isospecial  relativity}~[4],
characterized   by   infinitely     many   possible,      generalized,
positive--definite   units.  The selection  of the   special or general
relativity  is  then done via the  the {\sl assumed specific value of
the
unit},  the value   $\hat{I} \equiv I$
recovering  the   {\sl special}   relativity
identically, while the more general value $\hat{I} = \hat{I}_{gr}(x) =
[\hat{T}_{gr}(x)]^{-1}$  implies the selection of
the {\sl general} relativity (see below on for a broader dependence).

C)  The formulation permits the  achievement, also for the first time,
of a   unique   and  universal  symmetry of   gravitation,
Santilli's isopoincar\'{e} symmetry $\hat{P}$~[3--7]. In fact, as
pointed
our earlier, $\hat{P}(3.1)$ is the universal  symmetry of all infinitely
possible
line elements~(\ref{eq:two-5b}) which admit as  particular cases  the
Riemannian
forms. This resolves the historical difference between the general and
the special relativity whereby the  latter is  indeed equipped with  a
universal symmetry, the   Poincar\'{e} symmetry,  while  the former  is
not.  By  no means,  the   availability of  a  universal  symmetry for
gravitation is a mere  mathematical curiosity, because it carries  the
same rigid physical guidelines as provided for the special relativity,
which are lacking in the current formulation of gravitation.

D) The formulation permits {\sl the resolution  of known ambiguities in
the compatibility of the general and special relativites}. Recall that
the generators of any symmetry  are the total conserved quantities and
they remain   unchanged under isotopies~[4].  Then, the total
conservation laws of general relativity
in isominkoewskian formulation are established by mere visual
observation  of the  generators  of the   its  universal symmetry, the
isopoincar\'{e}   symmetry,  thus  avoiding    complex  and
controversial
calculations.  Moreover,  the  compatibility  of   relativistic    and
gravitational conservation laws  is   established by the mere   visual
observation   that these generators are the  same
for both relativities,  thus resolving additional known
controversies on  their   claimed lack of  compatibility,  and similar
occurrences  hold    for  other  aspects   (see~[4h]    for  the
preservation of weight at the  {\it relativistic} limit via  the use of
the
forgotten Freud identity). Therefore, the  lack of the rigid guidelines
of a  universal symmetry in gravitation  appears  to be the  origin of
some of the ambiguities here considered.

E)  The geometric unification  of the general and special relativities
constitutes  the foundation of a  novel   operator version of  gravity
studied in Sect. 7, which is as axiomatically consistent as relativistic
quantum  mechanics,   thus   avoiding known   problematic   aspects of
conventional forms of quantum gravity.

F) The formulation permits the introduction of {\sl a novel unification
of gravitational and electroweak  interactions based on the embedding
of the part truly  representing curvature, the gravitational unit
$\hat{I}_{gr}(x) = [\hat{T}_{gr}(x)]^{-1} > 0$,
in  the unit of unified  gauge theories}, whose studies were initiated
by Gasperini~[7o]  (see also the  review in  App.~A of
monograph~[6a]) and pointed out in Ref.~[4h].

G) The formulation permits {\sl  a novel relativistic and gravitational
treatment   of  antimatter at  both    classical  and quantum
levels}~[5]
studied in  the next section,  which avoids  the problematic
aspects   of the  Riemannian representation of antimatter outlined
below.

H)  The formulation permits {\sl a  direct geometric representation of
interior relativistic  and gravitational  problems}, for  instance,  a
direct representation of  the locally varying  speed of  light
$ c = c_0/n_4$ via the {\sl metric} of the  isominkowskian geometry with
$\hat{T}_{44} = n_4^{-2}\times g_{44}$. This latter
possibility is permitted by  the unrestricted functional dependence of
the   isometric which,    when  restricted to   a   sole
$x$--dependence,
characterizes {\sl exterior gravitational  problems in the homogeneous
and  isotropic  vacuum}, while the   use of a more  general dependence
characterizes      {\sl   interior   gravitational   problems   within
inhomogeneous and  anisotropic physical media} (with the understanding
that the background space remains homogeneous and isotopic).

The  {\sl axiomatic consistency} of  the isominkowskian formulation of
gravity is assured by the axiom--preserving character of the isotopies.

The {\sl  plausibility} of  the proposed   theory is illustrated  by  a
comparison  of properties A)--H) above  with the corresponding features
of the conventional formulation of gravity.

The {\sl experimental verification}  of the isominkowskian formulation
of gravity is  established   by the fact   that  Eq.s~(\ref{eq:two-14})
coincide
numerically  with  the conventional  Einstein  equations, as indicated
earlier. The isominkowskian formulation of gravity therefore possesses
the same verifications of the conventional Riemannian formulation.

\section{ Classical isodual isominkowskian  unification of the  special
and
general relativities for antimatter.}
\setcounter{equation}{0}

The current classical  representation of antimatter  is afflicted by a
number of  problematic aspects  at the   classical level,  as well  as
operator levels. The  mathematically correct  map between matter   and
antimatter   must    be  {\sl antiautomorphic}  (or,    more  generally,
anti--isomorphic),  as it  is  the case  for the  charge conjugation in
quantum mechanics.  The contemporary Riemannian  representation of
antimatter via
the simple change  of the sign of the  charge and magnetic  moments is
therefore insufficient.

Also, current theoretical   physics    admits  only {\sl one  type
of
quantization}, the conventional  one {\sl of matter},  and there  is no
separate quantization for antimatter. Therefore, the operator image of
the
current, classical, gravitational
representation of antimatter {\sl is not} the correct
charge conjugate {\sl antiparticle}, but merely  a {\sl particle} with
the
change of the sign of the charge and magnetic moments.

Moreover, the only energy--momentum tensor  available in the riemannian
geometry  is the conventional  one with positive--definite energy. Such
structure is manifestly incompatible   with the  {\sl negative--energy
}
solutions of relativistic field equations.

Because   of    these and  other problematic   aspects,   Santilli~[5]
introduced in  1985 a new  antiautomorphic map, called {\sl isoduality},
which can be applied to the entire formulations of matter, {\sl
beginning
at the  classical level}  and then continuing  at  the operator level.

The fundamental isodual map is that of the isounit

\begin{equation}
\hat{I} > 0 \rightarrow \hat{I}^d = -\hat{I}^\dagger = -\hat{I} < 0.
\label{eq:three-1}\end{equation}
This requires the reconstruction  of the entire isotopic formalism  in
such a way to admit $\hat{I}^d$, rather than $\hat{I}$,
 as the  correct left and right unit.

The most important quantities   are, again, new   numbers.   In  fact,
isoduality  must be  first  applied to  the  basic  isofields $\hat{F}
(\hat{a},+,\hat{\times})$ of
isoreals $\hat{a} = \hat{n}$,  isocomplex  $\hat{a} = \hat{c}$
   or isoquaternions  $\hat{a} = \hat{q}$,  yielding the
{\sl isodual isofields} $\hat{F}^d(\hat{a}^d,+,\hat{x}^d)$
(see~[3e] for comprehensive  studies), which
are rings of elements called {\sl isodual numbers}

\begin{equation}
\hat{a}^d = a^\dagger \times \hat{I}^d = -a^\dagger \times
\hat{I} = -\hat{a}^\dagger, \, (\hat{a}^d)^d \equiv \hat{a}.
\label{eq:three-2a}\end{equation}

\begin{equation}
\hat{n}^d = -n \times \hat{I}, \, \hat{c}^d = -\bar{c} \times \hat{I},
\,
\hat{q}^d = -q^\dagger \times \hat{I},
\label{eq:three-2b}\end{equation}
(where $\dagger$ denotes Hermitean conjugation and $\bar{}$ complex
conjugation), equipped with the isodual sum $ \hat{a}^d + \hat{b}^d =
(a + b)^\dagger \times \hat{I}^d$, and the {\sl isodual isoproduct}

\begin{equation}
\hat{a}^d\hat{\times}^d\hat{b}^d = \hat{a}^d \times \hat{T}^d \times
\hat{b}^d,
\, \hat{T}^d = -\hat{T}^\dagger = -\hat{T},
\label{eq:three-3}\end{equation}
under which $\hat{I}^d = (\hat{T}^d)^{-1}$ is the correct left and right
unit of $\hat{F}^d$,

\begin{equation}
\hat{I}^d\hat{\times}^d\hat{a}^d \equiv
\hat{a}^d\hat{\times}^d\hat{T}^d \equiv
\hat{a}^d, \, \forall \hat{a}^d \in \hat{F}^d,
\label{eq:three-4}\end{equation}
in which case  (only) $\hat{I}^d$ is
called the  {\sl isodual isounit}  and $\hat{T}^d$
called the {\sl isodual isotopic element}.

All  operations  of an  isofield are then   subjected to a simple, yet
significant isodual  map here  left  to the interested reader~[3e]. A
quantity is   called {\sl isoselfdual}   when    it is  invariant  under
isoduality. This  is the case for the  imaginary unit because $i^d =
-\bar{i}
\equiv i$  as well
as other quantities we shall identify later on.

A property of isodual isofields most important  for this paper is
that  {\sl isofields  and  isodual   isofields are  antiautomorphic
with
respect to each   others},   exactly as desired.   Another  important
property is that {\sl isodual isofields have a negative--definite norm}

\begin{equation}
\uparrow\hat{a}^d\uparrow^d = |a|\times\hat{I}^d = -|a|\times\hat{I} <0.
\label{eq:three-5}\end{equation}
As  a consequence, {\sl all physical   characteristics which are
positive
for  matter   become   negative    for  antimatter    under    isodual
representation},    as originally   assumed   at   the  discovery  of
antiparticles (Stueckelberg  and others).  This implies negative mass,
negative energy, negative (magnitude of  the) angular momentum, motion
back  ward in time,  change  of the sign   of charges and magnetic
moments, etc..

The  novelty is that  {\sl these negative characteristics are now
defined
with respect to negative units}, thus rendering inapplicable existing
arguments  against negative energy. As a  matter of fact, the referral
of the   negative--energy   solutions to  negative    units permits the
resolution of their un--physical behavior which historically motivated
the "hole theory" in second quantization~[5].

Similarly,  the isodual  representation renders  inapplicable existing
argument against motion backward in  time. In fact, motion backward in
time referred  to  a negative  unit of  time  is exactly as  causal as
motion forward in time referred to a  positive unit of time (for these
and other aspects, the interested reader may consult~[4h,5]).

The next  isodual  map must be applied   to the basic carrier  spaces,
yielding the {\sl isodual isominkowskian spaces}~[4d]

\begin{eqnarray}
\hat{M}^d(\hat{x}^d,\hat{\eta}^d,\hat{R}^d)&:&\, \hat{x}^d = -\hat{x},
\,
\eta^d = -\hat{\eta}, \hat{I}^d = -\hat{I} = (\hat{T}^d)^{-1} =
-\hat{T}^{-1} <0, \\
\nonumber \\
(\hat{x}^d - \hat{y}^d)^{\hat{2}d}& =& (\hat{x}^{\mu d} - \hat{y}^{\mu
d})
\hat{\times}^d \hat{N}^d_{\mu\nu}\hat{\times}^d(\hat{x}^{\nu d}
-\hat{y}^{\nu d}) = \nonumber \\
&=& [(x^\mu - y^\mu)\times \hat{\eta}^d_{\mu\nu} \times (x^\nu - y^\nu)]
\times \hat{I}^d = \nonumber \\
&=& [-(x^1 - y^1) \times T_{11}(x, ...) \times (x^1-y^1) - \nonumber \\
& & - (x^2-y^2) \times T_{22}(x, ...)(x^2 - y^2) - \nonumber \\
& &- (x^3 - y^3)T_{33}(x, ...) \times (x^3-y^3) + \nonumber \\
& & + (x^4 - y^4) \times T_{44}(x, ...) \times (x^4 - y^4)]
\times \hat{I}^d,\\
\nonumber \\
\hat{T}^d &=& Diag.(-T_{11}, -\hat{T}_{22}, -\hat{T}_{33},
-\hat{T}_{44}), \,
\hat{T}_{\mu\mu} > 0,
\label{eq:three-6b}\end{eqnarray}
with  all remaining properties  conjugated  with respect to
$\hat{M}(\hat{x},\hat{\eta},\hat{R})$  here
omitted for brevity.

The next isodual map is that of the  underlying calculus, yielding the
{\sl isodual differential calculus}~[3f] which is based on the rules
\begin{equation}
\hat{d}^d\hat{x}^{\mu d} = \hat{I}^{d\mu}_\nu \times d\hat{x}^{d\nu}
\equiv \hat{d}\hat{x}^\mu,
\hat{\partial}^d\hat{/}^d\hat{\partial}\hat{x}^{d\mu} =
-\hat{\partial}\hat{/}\hat{\partial}\hat{x}^\mu, etc.
\label{eq:three-7}\end{equation}
The next conjugation is that of the applicable geomety,
yielding the {\sl isodual
isominkowskian geometry}~[3e,4g] which  can  be constructed  via the
above rules and the following isodualities

\begin{equation}
\hat{\Gamma}^d_{\alpha\beta\gamma} = -\hat{\Gamma}_{\alpha\beta\gamma},
\,
\hat{D}^d\hat{X}^{d\mu} = \hat{D}\hat{X}^\mu, \, \hat{X}^{\beta d}_
{\uparrow\mu d}
 = -\hat{X}^{\beta}_{\uparrow\mu},
\label{eq:three-8b}\end{equation}
\begin{equation}
\hat{R}_\alpha^\beta{}_{\gamma\delta}^d = -\hat{R}_\alpha^\beta
{}_{\gamma\delta}, \, \hat{R}_{\mu\nu}^d = -\hat{R}_{\mu\nu}, \,
\hat{R}^d = - \hat{R}, etc,
\label{eq:three-8c}\end{equation}

The {\sl isodual isofield equations} then read

\begin{equation}
\hat{G}_{\mu\nu}^d = \hat{R}_{\mu\nu}^d -
\hat{\frac{1}{2}}^d\hat{\times}^d
\hat{N}_{\mu\nu}^d\hat{\times}^d\hat{R}^d = \hat{k}^d\hat{\times}^d
\hat{\tau}_{\mu\nu}^d,
\label{eq:three-14}\end{equation}
and they result to be the negative image  of Eq.s.~(\ref{eq:two-14}).
Other aspects are studied in the forthcoming paper [9].

A  property of the isodual  isominkowskian geometry most important for
this  paper    is  that  {\sl the   isodual   energy--momentum  tensor
is
negative--definite}, exactly as needed   for overall consistency in  any
theory of antimatter.

The next  isoduality is that  of  the basic isosymmetries  studied
in~[4d,4e], yielding the {\sl
isodual   isorotation, isodual isolorentz  and
the isodual isopoincar\'{e} symmetry}.  The next conjugation is that of
the isorelativity, yielding the {\sl isodual isospecial relativity}, and
of analytic  mechanics, yielding the {\sl isodual isoanalytic mechanics}
which are not outlined for brevity~[4d,4h].

In order   to  apply the above  results   to  a unified   treatment of
antimatter, the reader should be aware that all preceding
formulations admit as particular cases  the {\sl isodual numbers,
isodual
Minkowskian  geometry, isodual  Poincar\'{e}  symmetry} and   {\sl
isodual
special    relativity},  namely,  they     admit  hitherto    unknown
antiautomorphic  images  of  {\sl conventional}  theories  which  are
here
assumed for the {\sl relativistic} characterization of antimatter in
vacuum.

The {\sl classical isomincowskian reformulation of general relativity
for
antimatter }  is then  given by the {\sl isodual isospecial relativity}
under  the particular realization of  the isodual isotopic element and
isodual isounit

\begin{equation}
\hat{M}^d(\hat{x}^d,\hat{\eta}^d,\hat{R}^d): \, \hat{\eta}^d(x^d) =
\hat{T}^d_{gr}(x^d)\times\eta, \, \hat{I}_{gr}^d(x) =
[\hat{T}^d_{gr}(x^d)]
^{-1},
\label{eq:three-15}\end{equation}
which admits as particular case the {\sl isodual special relativity} for
$\hat{I}_{gr}^d = I^d = -\hat{I}$.

The  axiomatic   consistency   of   the  above   classical,   isodual,
relativistic and  gravitational representation of antimatter appears to
be  established   beyond  reasonable   doubts. Its     plausibility is
established  by    its resolution   of the  problematic     aspects of
conventional formulations   (see also next   sections).  Tshe  physical
validity  of the classical   isodual  theory  is  established by   the
verification of the sole  classical experimental data on antimatter
available at
this  writing, those   under {\sl electromagnetic}  interactions,
because
those under {\sl gravitational}  interactions are still unknown.

We  can therefore  conclude these introductory  considerations on the
unification  of the special  and  general relativity for antimatter by
indicating that the  emerging  novelty warrants additional studies  in
the  field.  In  fact, the  isodual  theory  permits  a mathematically
correct study of  the gravitational field  of astrophysical bodies made
up   of antimatter  and  the  initiation   of studies  for  the future
experimental resolution whether a  far away star  or quasar is made up
of matter or of antimatter [5].

\section{Problematic aspects of quantum gravity }
\setcounter{equation}{0}

We now pass to the study of the unification of the special and general
relativities at the {\sl operator} level. Again, we  have to insist that
it would  be unreasonable to expect a  comprehensive treatment in this
introductory paper of  all  operator  and field theoretical   aspects,
because they are so many to discourage even a partial outline.

The  remaining  scope of this  paper  is merely  that of  identify the
essential  operator   foundations  of the isospecial  relativity,  and
establish its plausibility as  compared to  other operator forms  of
gravity at the simplest possible level of "first quantization". In any
reasonable conduction of  research, quantum field  theoretical aspects
can  only be considered after  the identification  of hitherto unknown
isogravitational grounds in first   quantization. By keeping   in mind
that  the physical validity  of  the conventional quantum treatment  of
gravity is still  debated after three  quarters of a centuries  of
studies, the  reader should not expect  the resolution of the physical
validity of the  isooperator treatment of  gravity  in its first
presentation.

The basic open problems in  the operator version of general relativity
are the following:

A) On   one side,   relativistic   quantum  mechanics (RQM)  needs   a
meaningful   Hamiltonian    while, on  the   other   side,  Einstein's
gravitational in vacuum has a null Hamiltonian.

B) There is the need of an operator  gravity which is as axiomatically
consistent as the conventional RQM, i.e., invariant under its own time
evolution with physical quantities  which are  Hermitean--observable at
all times, with unique and invariant numerical predictions, etc..

C) Recent studies on  interior  gravitational problems of  black holes
(see, e. g., the studies  by Ellis et all.~[7s] and references quoted
therein)   have    indicated   that operator  gravity     should be  a
{\sl nonunitary}  image   of conventional  quantum theories,  as needed,
e. g., for a representation of irreversibility.

The isominkowskian reformulation of  general relativity permits  a new
{\sl operator version  of   gravity for matter}   here called {\sl
operator
isotopic gravity, or  operator isogravity} (OIG)  for short, which is
based on an axiom--preserving lifting  of the {\sl unit}  of RQM from
the
trivial value $I = Diag.(1, 1, 1, 1)$ to the gravitational value
$\hat{I}_{gr}(x)$. As such
OIG   requires  no Hamiltonian   at   all,  thus resolving  the  first
historical  problem. The axiomatic consistency  of the proposed OIG is
guaranteed  by the  preservation of the  abstract axioms  of  RQM, only
realized in a more general way, including form--invariance, Hermiticity
of observables at all times, etc.,  thus resolving the second problem.
Finally, the  proposed OIG  is a rather  natural nonunitary  image  of
conventional RQM, thus verifying the third condition.

The  isodualites  then permit  an  antiautomorphic {\sl isodual
operator
isogravity}  (IOIG)    which is    based  on   the  negative--definite
gravitational units of the preceding section.

A  preliminary    comparison of the   OIG  with  the conventional theory
currently used, called {\sl quantum gravity}  (QG)~[12] is in order.
Even
though    both  theories are  of  operator    character,  OIG and  the
conventional QG are {\sl inequivalent}, as illustrated by the fact that,
e.g., they are  defined on inequivalent  Hilbert spaces and fields. In
any case, the term "quantum" would be inappropriate under a nonunitary
structure as requested by condition C) and, for this and other reasons
we have preferred the generic term "operator".

As we shall see, the term  "quantum" is also questionable for "quantum
gravity"  because  this latter theory   too is outside the equivalence
classes of RQM.  In shot, this  paper presents evidence supporting the
fact that  a "quantum"  version of gravity,  that is  a version obeying
{\sl conventional}  quantum mechanics,   {\sl cannot}  exist, as one
can
anticipate from the  noncanonical structure of the Riemannian geometry
(Theorem 1 of Sect.1).

The identification  of the following  problematic aspects of  QM [3g,13]
is in
order, not only because necessary  to appraise the plausibility of OIG
on a comparative basis, but also because they do not appear to be well
known in the specialized literature in the field:

I) {\sl QG does not possess an invariant basic  unit}, as established by
theorem  1.    Also, the   time  evolution   of   QG  is   necesserily
{\sl nonunitary}  (otherwise QG   would be a    trivial element of   the
equivalence class of RQM),  thus confirming the  lack of invariance of
the basic unit,

\begin{equation}
I \rightarrow I' = U \times I \times U^\dagger \not = I.
\label{eq:four-1}\end{equation}
Therefore, {\sl all forms of QG gravity  with a nonunitary time
evolution
(hereon assumed)   lack   unambiguous   applications to   experimental
verifications};

II) {\sl QG does not preserve  Hermiticity in time}  when formulated on
a
conventional Hilbert space over a conventional field. In fact, under a
nonunitary transform, the familiar associative  modular action of  the
Schr\"{o}dinger's   representation  $H\times|\psi >$,   where $H$  is
an  operator
Hermitean at the initial time, becomes

\begin{eqnarray}
U \times H \times |\psi > &=& U \times H \times U^\dagger \times (U
\times
U^\dagger)^{-1} \times U \times |\psi> = \hat{H} \times \hat{T} \times |
\hat{\psi}>, \nonumber \\
U \times U^\dagger \not = I, \, & & \hat{T} = (U \times U^\dagger)^{-1},
\,
|\psi> = U \times |\psi>, \, \hat{H} = U \times H \times U^\dagger.
\label{eq:four-2}\end{eqnarray}
By nothing that $\hat{T}$ is Hermitean, $\hat{T} = (U\times
U^\dagger)^{-1}=
\hat{T}^\dagger$, the initial condition of Hermiticity of H,
$<\psi|\times {H\times|\psi>} = {<|\times H^\dagger}\times|\psi>$,
when applied to the {\sl conventional} Hilbert space $\cal H$ with
states
$|\hat{\psi} >, |\hat{\phi}>$, etc, requires the action of the
transformed
operator~(\ref{eq:two-2a}) on a {\sl conventional} inner product,
resulting in the
expressions

\begin{equation}
<\hat{\psi}|\times\{\hat{H}\times\hat{T}\times|\hat{\psi}>\} =
\{<\hat{\psi}|\times\hat{T}\times\hat{H}^\dagger\}\times|\hat{\psi}>,
{\rm i.e., }
\hat{H}^\dagger = \hat{T}^{-1}\times\hat{H}\times\hat{T}\not = \hat{H}.
\label{eq:four-3}\end{equation}
As  such, Hermiticity  is  not preserved under   the time evolution of
nonunitary theories when formulated   on conventional space $\cal H$
over
conventional fields C, because of the lack of general commutativity of
$\hat{T}$  and $\hat{H}$.
Consequently,  {\sl QG does not admit physically acceptable
observables}, an occurrence first indicated by Lopez~[13a].

III)    {\sl QG  does not   admit   invariant physical  laws  and
numerical
predictions}.   These are an  evident consequence of the nonunitarity
of the time evolution and do not require further elaboration.

An objectives of this  paper is to  see whether our OIG permits the
resolution of at least  some of the  above problematic aspects. At any
rate, the lack of resolution until now of the above problematic aspects
does   warrant the study of structurally  novel
operator formulations of gravity.

\section{Isotopic completion  of relativistic  quantum mechanics and
its
isodual}
\setcounter{equation}{0}

We now outline for the reader's convenience the {\sl operator
isotopies}, first  proposed in ref.~[3b]  of 1978 and then
studied  by numerous  authors and known under the name of
{\it relativistic hadronic mechanics} (RHM)..  This outline
is recommendable because
axiomatic maturity of the new mechanics has been reached oly   recently
in memoir [3g], following the achievement of mathematical maturity in
Ref. [3f].

The operator  isotopies are nowadays  defined as  {\sl maps  of any
given
linear,  local and unitary structure  into  its most general  possible
nonlinear, nonlocal and nonunitary forms, which are however capable of
restoring  linearity, locality  and  unitarity on suitable generalized
spaces over generalized fields}.

The  fundamental  isotopy  is    the   lifting of   the   conventional
$(3+1)$--dimensional unit of relativistic quantum mechanics (RQM) as for
classical isotheories

\begin{equation}
I \, \rightarrow \, \hat{I}(x, \dot{x}, \ddot{x}, \psi, \partial\psi,
\partial\partial\psi, ...) > 0,
\label{eq:five-1}\end{equation}
with the additional dependence on the  wavefunctions and their
derivatives.

Jointly the  conventional associative product  among generic operators
$A, B, A\times B = AB$
(e.g., the  elements of an enveloping  algebra $\xi$)  must be
lifted into the form

\begin{equation}
 A \times B \, \rightarrow \, A\hat{\times} B  = A \times \hat{T} \times
B,
\label{eq:five-2}\end{equation}
where $\hat{T}$ is fixed for all elements of $\xi$.
Under the conditional $\hat{I} = \hat{T}^{-1}, \hat{I}$
results to be the correct (left and right) generalized unit of the new
theory,

\begin{equation}
\hat{I} \hat{\times } A  = A\hat{\times} \hat{I} \equiv A, \: \forall A
\in \xi.
\label{eq:five-3}\end{equation}
The  lifting  $A\times B \rightarrow A\hat{\times} B$
is  called {\sl isotopic}   in the  sense that it
preserves the all original axioms, including associativity,
$A\hat{\times}
(B\hat{\times})C = (A\hat{\times}b)\hat{\times}C$.

The most direct way to  construct  operator isotopic methods is  {\sl
via
nonunitary transforms of}   RQM, called,   {\it isotopic completion   of
relativistic quantum mechanics} [3g]. Let U characterize  a
conventional nonunitary
transformation on a conventional Hilbert space $\it H$ over a
conventional
field C of  complex numbers. The deviation  of the transform from I is
assumed to be precisely  equal  to the  isounit of the  new theory,
$U\times U^\dagger = \hat{I} = \hat{I}^\dagger\not = I$.
For the case of the canonical commutation rules we have

\begin{equation}
U \times U^\dagger = \hat{I} = \hat{I}^\dagger \not = I, \; \hat{T} =
(U\times U^\dagger)^{-1} = \hat{I}^{-1} = \hat{T}^\dagger,
\nonumber \end{equation}
\begin{equation}
\bar{x}_\mu = U \times x_\mu\times U^\dagger, \; \bar{p}_\nu =
U\times p_\nu\times U^\dagger,
\end{equation}
\begin{equation}
U\times[x_\mu, p_\nu]\times U^\dagger =
 \bar{x}_\mu\times\hat{T}\times\bar{p}_\nu - \bar{p}_\nu\times\hat{T}
\times x_\mu  = \end{equation}
\begin{equation}
= i\times\eta_{\mu\nu}\times U\times I\times U^\dagger  =
i \times \eta_{\mu\nu}\times\hat{I},
\label{eq:five-4}\end{equation}

where one  should note that the isounit  and the isotopic element have
the correct Hermiticity   property, and the  emerging  new commutation
rules have precisely the needed isotopic character of RHM.

It is  easy to see that the  above isotopic theory suffers  of
essentially   the  same problematic     aspects  of QG    indicated in
Sect. 4. In  fact, the above theory   {\sl is not} form--invariant
under
additional nonunitary transforms  when treated  with the  conventional
mathematics  of   RQM. In particular,   nonunitary   transforms do not
preserve  the  original   unit     $\hat{I}$, thus   preventing
unambiguous
applications   to measurements.  Moreover,    such a  theory does  not
preserve Hermiticity at  all times,  thus preventing the   unambiguous
representation  of observable. Finally, it is  easy to see that such a
theory does not  possess  invariant numerical predictions,  because of
the lack of  invariance of  the  special functions needed  in the data
elaboration.

A resolution   of  these problem   requires the   construction of  the
isotopies of the entire structure of RQM~[3f,3g] without any exception
known to  this author.  In  fact, the characterization  of RHM via the
use in part of isotopic structures and  in part of conventional quantum
structures, is afflicted by  rather fundamental  inconsistencies which
often remain undetected by the non--expert in  the field.  Equivalently
we  can  say  that   the  problematic aspects    are  resolved  by the
application of the  nonunitary  transform to the  {\sl totality}  of the
conventional formalism, including numbers,  fields, metric and Hilbert
spaces, algebras, geometries, etc.

In this way,
RQM, RHM is characterized by the following primary structures:

{\bf A)}  the lifting of the field  $C = C(c, +, \times)$
of complex numbers into the
isofields $\hat{C}( \hat{c}, +, \hat{\times})$
of isocomplex numbers of  Sect.2 with the isofield of
isoreal numbers $\hat{R}( \hat{n}, +, \hat{\times})$  as a particular
case.

{\bf B)}  the lifting of  the conventional Minkowski
space  $M = M( x, \eta, R)$
 into the {\sl iso\-min\-kows\-ki\-an  space} of Sect. 2
plus  the enlarged
functional dependence of isotopic element and, thus of the isometric,

\begin{equation}
\hat{M}(\hat{x},\hat{\eta},\hat{R}): \; \hat{x} = x \times \hat{I},
\hat{\eta} = \hat{T}(x, \dot{x}, \ddot{x}, \psi, \partial\psi,
\partial\partial\psi, ...)\times \eta, \hat{I} = \hat{T}^{-1}.
\end{equation}

{\bf C)} the lifting  of the conventional  Hilbert space $\cal H$
 with  states $|\psi>, |\phi>, ...$
and inner product $<\phi|\psi>\in C(c,+,\times)$
into the   {\sl isohilbert}  space $\hat {\cal H}$ with {\sl isostates }
$|\hat{\psi}>, |\hat{\phi}>, ...,$
{\sl isoinner product and isorenormalization}

\begin{equation}
<\hat{\phi}\uparrow\hat{\psi}> =
<\hat{\phi}|\times\hat{T}\times|\hat{\psi}>
\times\hat{I}\in \hat{C}(\hat{c},+,\hat {\times}),
\label{eq:five-6a}\end{equation}
\begin{equation}
<\hat{\psi}|\times\hat{T}\times|\hat{\psi}> = I,
\label{eq:five-6b}\end{equation}

{\bf D)} the lifting of theory of linear operators in Hilbert space into
the
corresponding theory  on $\hat {\cal H}$,  including the  lifting  of
the familiar
eigenvalue equations $H\times|\psi> = E_0\times|\psi>$
 into the {\sl isoschr\"{o}dinger equation}

\begin{equation}
H\hat{\times}|\hat{\psi}> = H \times \hat{T}\times|\hat{\psi}> =
\hat{E}\hat{\times}|\hat{\psi}> =
E\times\hat{I}\times\hat{T}\times|\hat{\psi}
> \equiv E\times|\hat{\psi}, \;\; E\not = E_0,
\label{eq:five-7}\end{equation}
indicating that the final numbers  of the theory are conventional; the
proof that the isoeigenvalues of  an isohermitean operators area real,
etc.

{\bf E)} the {\sl isodifferential} calculus on  $\hat{M}(\hat{x},
\hat{\eta}, \hat{R})$ of Sect. 2,

\begin{equation}
\hat{d}x^\mu = \hat{I}^\mu_\nu\times dx^\nu, \: \hat{\partial}_\mu =
\hat{\partial}/\hat{\partial}x^\mu = \hat{T}_\mu^\nu\times\partial_\nu =
\hat{T}_\mu^\nu\times\partial/\partial x^\nu,
\label{eq:five-8a}\end{equation}
\begin{equation}
\hat{\partial}x^\mu/\hat{\partial}x^\nu = \delta^\mu_\nu, \;
\hat{\partial}
x_\mu/\hat{\partial}x^\nu = \hat{\eta}_{\mu\alpha} \:\:
\hat{\partial}x^\alpha
/\hat{\partial}x^\nu = \hat{\eta}_{\mu\nu},
\label{eq:five-8b}\end{equation}
\begin{equation}
\hat{\partial}x^\mu/\hat {partial}\hat{x}_\nu =
\hat{\eta}^{\mu\alpha}\times\hat{\partial}x_
\alpha/\hat{\partial}\hat{x}_\nu = \hat{\eta}^{\mu\nu};
\label{eq:five-8c}\end{equation}

{\bf F)} the lifting of Heisenberg's  equation, the  equation on the
linear
momentum and the   fundamental commutation  rules  into the  following
{\sl isoheisenberg equation,    isolinear   momentum} and  {\sl
fundamental
isocommutation rules}, first formulated in an axiomatically complete and
correct form in memoir [3g]

\begin{equation}
i \times \hat{d}/\hat{d}\hat{t} = [\hat{A}, \hat{}\hat{H}] = \hat{A}
\hat{\times}\hat{H} - \hat{H}\hat{\times}\hat{A},
\label{eq:five-9a}\end{equation}
\begin{equation}
p_\mu\hat{\times}|\hat{\psi}>=p_\mu\times\hat{T}\times|\hat{\psi}>
=-i \times \hat{\partial}_\mu|\hat{\psi}>=-i \times
\hat{T}_\mu^\nu \times \partial_\nu|\hat{\psi}>,
\label{eq:five-9b}\end{equation}
\begin{equation}
[x_\mu,\hat{p}_\nu]\hat{\times}|\hat{\psi}>=(\hat{x}_\mu\times\hat{T}\times
\hat{p}_{nu}-\hat{p}_{nu}\times\hat{T}\times\hat{x}_\mu)\times
\hat{T}\times|\hat{\psi}>=i \times \hat{\eta}_{\mu\nu}\times|\hat{\psi},
\label{eq:five-9c}\end{equation}

{\bf G)} the  lifting of expectation  values  $<A> = <\psi|\times
A\times|
\psi>/<\psi|\psi>$ into the {\sl isoexpectation values}

\begin{equation}
\hat{<}A\hat{>} = <\hat{\psi}|\times\hat{T}\times A\times\hat{T}\times|
\hat{\psi}>/<\hat{\psi}|\times\hat{T}\times|\hat{\psi}>;
\label{eq:five-10}\end{equation}

and the compatible liftings of the remaining aspects of RQM~[3g].

{\bf H) } The preceding formalism is completed   with the {\sl isotopies
of the
naive  or  symplectic quantization}, which  establish  the unique and
unambiguous derivability of RHM from  the corresponding classical
isoanalitic
mechanics (Sect. 2) (see~[3f,4h] for brevity).

The following comments are in order. First, it is easy to see that the
theory is highly nonlinear (hereon referred to a nonlinearity {\sl in
the
wavefunction}), e.g.,   isoeigenvalue equations~(\ref{eq:five-7}) can
be written
explicitly

\begin{equation}
\hat{H}\hat{\times}|\hat{\psi}>= \hat{H}\times\hat{T}(x, p, \psi,
\partial\psi, ...)\times|\hat{\psi}>= E\times|\hat{\psi}.
\label{eq:five-11}\end{equation}
Nevertheless, the theory does satisfy  the conditions of linearity  in
isospace,

\begin{equation}
\hat{A}\hat{\times}(\hat{a}\hat{\times}\hat{x}+\hat{b}\hat{\times}\hat{y})
=
\hat{a}\hat{\times}\hat{A}\hat{\times}\hat{x} +
\hat{b}\hat{times}\hat{A} \hat {\times}
\hat{y}, \; \hat{A} \in \xi, \; \hat{a},\hat{b} \in
\hat{R}(\hat{n},+,\hat{x}),
\; \hat{x}, \hat{y} \in \hat{M}(\hat{x},\hat{\eta},\hat{R}).
\label{eq:five-12}\end{equation}
and it is called {\sl isolinear}.

Also, all nonlinear theories can be  {\sl identically} reformulated in a
isolinear form   in which  all nonlinear   terms are  embedded  in the
isotopic element, e.g.,

\begin{equation}
H(x, p, \psi, ...)\times|\psi>\equiv H_0(x, p)\times\hat{T}(\psi, ...)
\times|\psi>= H_0\hat{\times}|y>.
\label{eq:five-13}\end{equation}
This  resolves the  loss  of  the  superposition   principle which  is
inherent  in  all  nonlinear   theories, with  consequential  loss  of
consistent treatment of composite systems [3g,13].

Second,   the theory verifies the  conditions  of locality in isospace
over    isofields  whenever all  nonlocal terms   are  embedded in the
isounit,  and it is then  called   {\sl isolocal}. Also, {\sl all
nonlocal
theories can be identically rewritten in an isolocal form}.

Finally, the theory readily reconstructs unitarity in isospace, and it
is called {\sl isounitary}.  In fact, nonunitary  transforms of the same
"magnitude" $\hat{I}$ (i.e., such that $W\times W^\dagger = \hat{I}$)
can always be written [3g]

\begin{equation}
W = \hat{W}\times\hat{T}^{\frac{1}{2}},
\label{eq:five-14}\end{equation}
and be, therefore, turned into the {\sl
isounitary transforms} on $\hat {\cal H}$,

\begin{equation}
\hat{W}\hat{\times}\hat{W}^{\hat{\dagger}} =
\hat{U}^{\hat{\dagger}}\hat{\times}
\hat{U} = \hat{I},
\label{eq:five-15}\end{equation}
Thus, {\sl all   nonunitary theories can  be identically  rewritten in
an
isounitary form}.

As  an incidental  comment, one should  note that  {\sl the admission
of
nonunitary transforms with "magnitude"   $U\times U^\dagger = \hat{I}$
 different than  $\hat{I}$  would
imply the  transition  to   a    different physical   system}.   The
transformation theory of RHM  is therefore restricted for each  system
considered   to  the selected  $\hat{I}$,  {\sl in   exactly  the  same
way as
conventional RQM  restricts   the admitted transforms   to  those with
conventional "magnitude" $I$ only}.

In view of  the above properties, RHM  is  form--invariant and resolves
the   physical problematic   aspects   of other   nonunitary  theories
indicated in Sect. 2. In fact, we have the following properties:

i){\sl RHM possesses an  invariant isounit}. In fact,$ \hat{I}$
is numerically  preserved
under isounitary transforms and it is preserved in time,

\begin{equation}
\hat{I}\rightarrow\hat{I}' =
\hat{U}\hat{\times}\hat{I}\hat{\times}\hat{U}^{
\hat{\dagger}}\equiv \hat{I},
\label{eq:five-16a}\end{equation}
\begin{equation}
i \times d\hat{I}/dt = [\hat{I},\hat{H}] = \hat{I}\hat{\times}H -
H\hat{\times}
\hat{I} = H - H \equiv 0;
\label{eq:five-16b}\end{equation}
with   consequential  unambiguous   application  of    the  theory  to
measurements;

ii) {\sl RHM  preserves Hermiticity and observability  at all times}. In
fact, the condition of Hermiticity on $\hat{\cal H}$ over
$\hat{C}(\hat{c},+,
\hat{\times})$ now reads

\begin{equation}
\{<\hat{\psi}|\times\hat{T}\times
H^{\hat{\dagger}}\}\times|\hat{\psi}>=<
\hat{\psi}|\times\{H\times\hat{T}\times|\hat{\psi}>\},
\label{eq:five-17a}\end{equation}
\begin{equation}
H^{\hat{\dagger}}\equiv\hat{T}^{-1}\times\hat{T}\times
H^\dagger\times\hat{T}^{-1}\times\hat{T}\equiv H^\dagger = H,
\label{eq:five-17b}\end{equation}
and, as   such,  it {\sl coincides}  with  the  Hermiticity on   $\cal
H$ over
$C(c,+\times)$.
Therefore, {\sl all observables of RQM remain observables for RHM}.

iii) {\sl RHM  possesses invariant numerical  predictions, physical laws
and special functions}.  This is due to the invariance of the isounit
and
of the isoassociative product,
$\hat{U}\hat{\times}(A\hat{\times}B)\hat{\times}
\hat{U}^\dagger = \bar{A}\hat{\times}\bar{B}$;
the  invariance of the fundamental isocommutation rules,

\[
\hat{U}\hat{\times}(\hat{x}_\mu\hat{\times}\hat{p}_\nu -
\hat{p}_\nu\hat{\times}\hat{x}_\mu)\hat{\times}\hat{U}^\dagger\hat{
\times}|\hat{\psi}>=(\bar{x}_\mu\hat{\times}\bar{p}_\nu-\bar{p}_\nu\hat{\times}
\bar{x}_\mu)\hat{\times}|\bar{\psi}>=i \times
\hat{\eta}_{\mu\nu}\times|\hat{\psi}>,
\]
\begin{equation}
|\bar{\psi}>=\hat{U}\hat{\times}|\hat{\psi}>, \; \; \bar{x}_\mu =
\hat{U}
\hat{\times}\hat{x}_\mu\hat{\times}\hat{U}^\dagger, \; \;
\bar{p}_{nu} =
\hat{U}\hat{\times}\hat{p}_{nu}\hat{\times}\hat{U}^\dagger.
\label{eq:five-18}\end{equation}
and other aspects (see~[3h,3g] for all details).

Even though evidently not unique, RHM is {\sl directly universal} in the
sense  of admitting all infinitely  possible, well behaved, nonlinear,
integro--differential    signature--preserving deformations  of      the
Minkowski metric  $\hat{\eta} = \hat{T}\times\eta$
 (universality)   directly  in the frame   of the
observer (direct universality).

We also point out  that RHM is a   "completion" of RQM much  along the
celebrated argument by Einstein, Podolsky and Rosen~[14a] and, for this
reason,  it  is also called  {\sl isotopic  completion of RQM}. In fact,
Eq.s~(\ref{eq:five-7}) constitute   an explicit and  concrete
realization   of the
theory  of "hidden variables"  $\lambda$~[14b] actually realized as
"hidden
operators", $\lambda = \lambda(x, \dot{x}, \psi, \partial\psi, ...) =
\hat{T}$.
  It  should  be  indicated  that the  celebrated   von
Neumann theorem~[14c] and Bell's inequalities~[14d] {\sl do not apply}
to
RHM, trivially, because   of its essential  nonunitary structure
(see~[4h],
App.4.C for details).

The  mechanics is   called    "hadronic" because it   was   originally
recommended for the study of the structure and interactions of
hadrons~[3b]
with nonlinear, nonlocal and nonunitary internal effects, as well
as for all  interactions of particles  with appreciable overlapping of
the wavepackets, irrespective of whether the charges are point--like or
not. The application to hadrons remains the main objective of RHM, as
poutlined in Web Site [7t], Page 19, Sect. V.

In this paper  we shall study  the particularization  of  RHM for OIG,
thus opening intriguing possible  relationships between what are today
called  "strong  interactions" and gravitation planned    for study
elsewhere. At this point we merely  note that all verifications of
RHM~[4h,4i,5d] may eventually result to be verifications of OIG because
of
the admission of the latter as a particular case.

The {\sl isodual   relativistic hadronic  mechanics}   (IRHM) is the
antiautomorphic image of RHM characterized  by the application of  the
isodual map to each and every quantity and operator, including:

a) the {\sl isodual isofields} $\hat{C}^d(\hat{c}^d,  +,
\hat{\times}^d)$
and $\hat{R}^d(\hat{n}^d, +, \hat{\times}^d)$
  with fundamental unit, elements   and product

\begin{equation}
\hat{I}^d = -\hat{I} = (\hat{T}^d)^{-1}<0, \; \;
\hat{c}^d = c\times\hat{I}^d, \; \; \hat{n}^d = n\times\hat{I}^d, \; \;
\hat{a}\hat{\times}^d\hat{b} = \hat{a}\times\hat{T}^d\times\hat{b},
\label{eq:five-19}\end{equation}

b) the {\sl isodual isominkowski space} $\hat{M}^d(\hat{x}^d,
\hat{\eta}^d,
\hat{R}^d)$ of Sect. 3;

c) the {\sl isodual isohilbert space} $\hat{\cal H}^d$
characterized by the following
isodual   isostates,     isodual  isoinner   product    and    isodual
isonormalization

\begin{equation}
|\hat{\psi}>^d = -|\hat{\psi}>^\dagger = -<\hat{\psi}|, \; \;
{}^d<\hat{\phi}|\times\hat{T}^d\times|\hat{\psi}>^d\times\hat{I}^d \in
\hat{C}^d, \; {}^d<\hat{\phi}|\times\hat{T}^d\times|\hat{\psi}>^d = 1,
\label{eq:five-20}\end{equation}

d)  the {\sl isodual isoassociative  operator algebra}  characterized by
the unit, elements and product

\begin{equation}
\hat{\xi}^d\; : \: \hat{I}^d, \;\;\; \hat{X}^d = -\hat{X}, \;\;
\hat{X}_i^d
\hat{\times}^d\hat{X}_j^d = \hat{X}_i\times\hat{T}^d\hat{X}^j,
\label{eq:five-21}\end{equation}

e) the {\sl isodual isoeigenvalues equations}

\begin{equation}
\hat{H}^d\hat{\times}^d|\hat{\psi}>^d
=\hat{E}^d\hat{\times}^d|\hat{\psi}>^d = E^d\times|\hat{\psi}>^d,
\label{eq:five-22}\end{equation}
with the correct {\sl negative} eigenvalues $E^d = -E$; and

f) the {\sl isodual dynamical equations}

\begin{equation}
i^d {\times}^d \times \hat{\partial}_t^d|\hat{\psi}>^d =
\hat{H}^d\hat{\times}^d|\hat{\psi}>^d =
E^d\times|\hat{\psi}>^d,
\label{eq:five-23a}\end{equation}
\begin{equation}
i^d {\times}^d \hat{d}^d\hat{A}^d/^ddt^d = [\hat{A},\hat{}\hat{H}]^d =
\hat{A}^d\hat{\times}^d\hat{H}^d - \hat{H}^d\hat{\times}^d\hat{A}^d,
\label{eq:five-23b}\end{equation}
\begin{equation}
\hat{p}_k^d\hat{\times}^d|\hat{\psi}>^d = - i^d {\times}^d
\hat{partial}^d_k|\hat{\psi}>^d,
\label{eq:five-23c}\end{equation}
\begin{equation}
[p_i, \hat{} r^j]^d = -\delta_i^j, \: [p_i, \hat{} p_j]^d = [r^i, \hat{}
r_j]^d
= 0,
\label{eq:five-23d}\end{equation}
where we have used the isoselfduality of the imaginary unit (Sect. 3)

g)  the  {\sl isodual  naive   or  symplectic  isoquantization},   which
establishes the unique derivability  of  the preceding formalism  from
the isodual isoanalytic mechanics of Sect. 3 (see~[3f,3g,4h] for
brevity).

The prof  of the equivalence of isoduality  and  charge conjugation is
presented in ref.~[5a].     Note that the    above isodual relativistic
hadronic  mechanics   admits  as   a  particular  case  the  {\sl
isodual
relativistic quantum   mechanics}.  The mechanics   emerging from our
study  and   their  unique  interconnecting   maps can   therefore  be
summarized as follows:

\begin{figure}
\setlength{\unitlength}{1mm}
\begin{picture}(160,70)
\put(30,65){\makebox(0,0)[bl]{MATTER}}
\put(120,65){\makebox(0,0)[bl]{ANTIMATTER }}
\put(25,50){\framebox(40,12)[tc]{Classical isoanalytic }}
\put(34,53){\makebox(0,0)[bl]{mechanics}}
\put(115,50){\framebox(40,12)[tc]{Classical isodual}}
\put(114,53){\makebox(0,0)[bl]{ isoanalytic mechanics}}
\put(25,10){\framebox(40,12)[tc]{Relativistic hadronic}}
\put(28,13){\makebox(0,0)[bl]{hadronic mechanics}}
\put(115,10){\framebox(40,12)[tc]{Isodual relativistich}}
\put(118,13){\makebox(0,0)[bl]{hadronic mechanics}}
\put(5,35){\makebox(0,0)[bl]{Isoquantiz.}}
\put(140,30){\makebox(0,0)[bl]{Isodual isoquantiz.}}
\put(80,52){\makebox(0,0)[bl]{Isoduality}}
\put(80,12){\makebox(0,0)[bl]{Isoduality}}
\thicklines
\put(65,56){\vector(1,0){50}}
\put(115,56){\vector(-1,0){50}}
\put(65,16){\vector(1,0){50}}
\put(115,16){\vector(-1,0){50}}
\put(45,50){\vector(0,-1){28}}
\put(135,50){\vector(0,-1){28}}
\end{picture}
\end{figure}

The most convincing evidence in favor of the isodual representation of
antiparticles can be seen  in  the structure of the   {\sl conventional}
Dirac equation  (see Sect. 7).   Here we  mention that   the  axiomatic
consistency of both,  the   isotopies and isodualities, can  be   also
established via the      following  {\sl new invariance  laws   of
the
conventional  inner   product of    Hilbert spaces}  presented
apparently for the first time in Ref. [3g]:

1) The  {\sl isoselfscalarity}, expressing    the invariance  under  the
charge  of the unit, here  expressed for isounit  independent from the
integration variables

\begin{equation}
<\psi|\times|\psi> = <\psi|\times\hat{T}\times|\psi>\times\hat{T}^{-1} =
<\psi\uparrow\psi> \in \hat{C},
\label{eq:five-24}\end{equation}

2) the {\sl isoselfduality}, which expressed the  invariance of the same
inner product under the antiautomorphic isodual map

\begin{equation}
<\psi|\times\hat{T}\times|\psi>\times\hat{I}\equiv {}^d<\psi|
\times \hat{T}^d \times |\hat{\psi}>^d\times\hat{I}^d.
\label{eq:five-25}\end{equation}

Evidently invariance~(\ref{eq:five-24}) expresses the
preservation of the abstract
quantum   axioms  under changes  of   the basic  unit  (isotopy), thus
establishing the  transition  from   RQM to RHM.
Invariance~(\ref{eq:five-24})
establishes that  the   same laws for  particles  are  also valid  for
antiparticles under their antiautomorphic interpretation (isoduality).

In summary, the isominkowskian formalism of Sect. 2 does indeed admit a
unique  and   unambiguous  operator  counterpart  which  {\sl cannot be}
"quantum", but   which is  nevertheless characterized by   the abstract
quantum axioms only in a more general realization.

The  isodual  isominkowskian formalism of  Sect. 3  does indeed admit a
unique operator counterpart which is the  antiautomorphic image of the
preceding one, thus being particularly suited for the representation of
antiparticles. Also,  the isodual  operator formalism originates  from
{\sl a new  quantization  specifically  built   for the  quantization
of
antimatter  into antiparticles}~[3g].  This  appears to resolve  the
long
standing impasse of  the  theoretical representation of  antimatter
caused by   the uniqueness of quantization {\sl vis--a--vis}
the duality  of matter and  antimatter, and opens  up a new horizon of
possibilities.

It should be stressed again that we have indicated in this section the
axiomatic consistency   and  plausibility   of  the  operator isotopic
treatment  of particles (generally  intended for particles in interior
conditions) and the isodual  isotopic representation of  antiparticles
(generally intended for antiparticles in interior conditions).

Preliminary studies indicate encouraging possibilities of experimental
verifications  under  the  conditions in  which   the formulations are
applicable, although such physical validity can only be established as
a result of collegian and predictably protracted investigations.

\section{ The universal isopoincar\'{e} symmetry and its isodual}
\setcounter{equation}{0}

Any appraisal  of OIG requires  at  least a  minimal knowledge of  the
operator  form of  its  universal symmetry,  the  {\sl isopoincar\'{e}
symmetry}~[4], specifically realized for  gravity, which is studied  in
this section following the achievement of sufficient maturity in
isosymmetries
by Kadeisvili in memoir [7a].

The isopoincar\'{e}  symmetry (also called in  the literature the {\sl
Santilli's  isopoincar\'{e} symmetry}~[4,5])  is the universal symmetry
of isoline element~(\ref{eq:two-5b}).  Therefore, the basic invariant
quantity is
not (length)$^2$, but the  broader structure
(length)$^2\times$(unit)$^2$.

The  isopoincar\'{e} symmetry can  be constructed via the step--by--step
application of the {\sl isotopies  of enveloping associative algebras,
Lie algebras, Lie groups,  transformation and representation  theory},
etc    called   Lie--Santilli  isotheory~[6--7]   and  consists  in the
reconstruction   of   the conventional  symmetry P(3.1) for the
generalized unit  $\hat{I} = \hat{T}^{-1}$.
Since $\hat{I}>0$,  one can see  form the inception that
$\hat{P}(3.1) \approx P(3.1)$.

Evidently we  cannot review  here the   rigorous  construction of  the
isosymmetry $\hat{P}(3.1)$ and we have to
limit ourselves for brevity to identify
its essential aspects.  The lifting  $P(3.1)\rightarrow\hat{P}(3.1)$
 is constructed  by
preserving the conventional generators of the Poincar\'{e} symmetry

\begin{equation}
X = \{X_k\} = \{M_{\mu\nu}, \times p_\alpha\}, \;\;\; M_{\mu\nu} = x_\mu
\times p_\nu - x_\nu\times p_\mu,
\label{eq:six-1a}\end{equation}
\begin{equation}
k = 1, 2, ..., 10, \;\;\;\; \mu,\nu = 1, 2, 3, 4,
\label{eq:six-1b}\end{equation}
and the conventional parameters

\begin{equation}
w = {w_k} = {(\theta, v), a} \in R,
\label{eq:six-2}\end{equation}
although they are now formulated in isospaces over isofiuelds,
and by submitting to an isotopies the {\sl operations} constructed on
them.

In  fact the above  quantities represent physical characteristics such
as energy, linear momentum,   angles, velocities, etc., which  are not
affected by short range interactions,  the latter being represented by
{\sl generalized operations among conventional physical quantities}.

The connected component of the  isopoincar\'{e} symmetry is given  by
$\hat{P}_0 (3.1) = S\hat{O} (3.1) \hat{\times} \hat{T} (3.1)$,
 where $S\hat{O} (3.1)$ is the  {\sl  connected iso\-lo\-rentz group}
first
introduced in~[4a] and $\hat{T}(3.1)$
is  the group of {\sl iso\-trans\-la\-ti\-ons}~[3d],
and it is characterized by the isotransforms on $\hat{M}(\hat{x},
\hat{\eta},
 \hat{R})$,

\begin{equation}
x' = \hat{A}(\hat{w})\hat{\times} x =
\hat{A}(\hat{w})\times\hat{T}\times x =
\tilde{A}(w)\times x, \;\;\; \hat{A} = \tilde{A}\times\hat{I},
\label{eq:six-3}\end{equation}
where the first form is the mathematically  correct one, the last form
being used for computational simplicity.

The   above   isotransforms can  be     expressed  via the  {\sl
isoexponention} in $\hat{\xi}$

\begin{equation}
\hat{A}(\hat{w}) = \hat{e}^{i\times X\times w} = \hat{I} +
(i \times X\times w)/1! +
(i\times  X\times w)\hat{\times}(i \times X\times w)/2! + ...
\label{eq:six-4}\end{equation}
characterized  by the {\sl isotopic Poincar\'{e}--Birkhoff--Witt
Theorem}~[3a,3d]  and   reducible  to  the  conventional
exponentiation  for
computational simplicity

\begin{equation}
\hat{A}(\hat{w}) = \hat{e}^{i\times X\times w} = \{e^{i\times X\times
T\times w}\}\times
\hat{I} = \tilde{A}(w)\times\hat{I}.
\label{eq:six-5}\end{equation}

The (connected component of the) {\sl isopoincar\'{e} group} can
therefore
be written as (or defined by )~[4]

\begin{equation}
\hat{P}_0{\rm (3.1)}\: :\: \hat{A}(\hat{w}) = \Pi_k\hat{e}^{i\times
X\times w} =
(\Pi_k e^{i\times X\times\hat{T}\times w})\hat{I} =
\tilde{A}(w)\times\hat{I},
\label{eq:six-6}\end{equation}

The preservation of the original dimension  is ensured by the
{\sl isotopic  Baker--Campbell--Hausdorff  Theorem~[3a,3d].  It  is easy
to see  that   structure~(\ref{eq:six-6})  forms   a    connected
Lie--Santilli
transformation isogroup} with {\sl isogroup laws}

\begin{equation}
\hat{A}(\hat{w})\hat{\times}\hat{A}(\hat{w}') =
\hat{A}(\hat{w}')\hat{\times}
\hat{A}(\hat{w}) = \hat{A}(\hat{w}+\hat{w}'), \;\;
\hat{A}(\hat{w})\hat{\times}
\hat{A}(-\hat{w}) = \hat{A}(0) = \hat{I} = \hat{T}^{-1}.
\label{eq:six-7}\end{equation}
Note that the use of the original Poincar\'{e} transform $x' =
A(w)\times x$
 would now
violate linearity in isospace,    besides  not yielding  the   desired
symmetry of isoseparation~(\ref{eq:two-5b}).

The isotopy   of  the discrete transforms    is  elementary~[4e],  and
reducible to the forms

\begin{equation}
\hat{\pi}\hat{\times}x = \pi\times x = (-r, x^4), \;\;
\hat{\tau}\hat{\times}
x = \tau\times x = (r, -x^4),
\label{eq:six-8}\end{equation}
where $\hat{\pi} = \pi\times\hat{I}$, and $\pi, \tau$ are
the conventional inversion operators.

To   identify  the  isoalgebra $\hat{p}_0(3.1)$ of $\hat{P}_0(3.1)$
we   use   the  {\sl isodifferential
calculus}~[3f]   on  $\hat{M}$   outlined earlier   which yields    the
{\sl
isocommutation rules}~[4]

\begin{equation}
[\hat{M}_{\mu\nu}, \hat{}\hat{M}_{\alpha\beta}] =
i\times (\hat{\eta}_{\nu\alpha}\times \hat{M}_
{\mu\beta} - \hat{\eta}_{\mu\alpha}\times \hat{M}_{\nu\beta} -
\hat{\eta}_{\nu\beta}\times \hat{M}_{\mu\alpha} +
\hat{\eta}_{\mu\beta}\times \hat{M}_{\alpha\nu}),
\label{eq:six-9a}\end{equation}
\begin{equation}
[\hat{M}_{\mu\nu}, \hat{}\hat{p}_\alpha] =
i\times (\hat{\eta}_{\mu\alpha\times }\hat{p}_\nu-
\hat{\eta}_{\nu\alpha}\times \hat{p}_\mu),
\label{eq:six-9b}\end{equation}
\begin{equation}
[\hat{p}_\alpha,\hat{}\hat{p}_\beta] = 0,
\label{eq:six-9c}\end{equation}
where $[A,\hat{}B]$ is the {\sl Lie--Santilli product} which satisfies
the Lie axioms in isospace, as one can verify.

The {\it isocasimir invariants} are given by

\[
C^{(0)} = \hat{I}(x, \dot{x}, \psi, \partial\psi, ...) = \hat{T}^{-1},
\]
\[
C^{(1)} = \hat{p}^{\hat{2}} = \hat{p}_\mu\hat{\times}\hat{p}^\mu =
\hat{\eta}^
{\mu\nu}\hat{p}_\mu\hat{\times}\hat{p}_\nu,
\]
\begin{equation}
C^{(3)} = \hat{W}_\mu\hat{\times}\hat{W}^\mu, \;\;\; \hat{W}_\mu = \in
{}_{\mu\alpha\beta\rho}\hat{M}^{\alpha\beta}\hat{\times}\hat{p}^\rho.
\label{eq:six-10}\end{equation}

The local isomorphism $\hat{p}_0(3.1)\approx p_0(3.1)$
is ensured by the positive--definiteness  of
$\hat{T}$.  Alternatively,  the use of the  generators in the form
$M^\mu_\nu = x^\mu\times p_\nu - x^\nu\times p_\mu$ yields
the {\sl conventional}  structure constants under a {\sl  generalized}
Lie product,  as one  can   verify. The   above local  isomorphism  is
sufficient, per s\'{e}, to guarantee the axiomatic consistency of RHM.

The  space components  $S\hat{O}(3)$, called {\sl   isorotations}    and
first
introduced in~[4b],  can be computed  from
isoexpontiations~(\ref{eq:six-6}) and
the space components $\hat{T}_{kk}$ of the isotopic element in diagonal
form,
$\hat{T} = Diag.(T_{\mu\mu})$,
yielding the explicit form in the (x,y)--plane

\[
x' =
x\times\cos(\hat{T}_{11}^{\frac{1}{2}}\times\hat{T}_{22}^{\frac{1}{2}}
\times\theta_3) -
\hat{y}\times\hat{T}_{11}^{-\frac{1}{2}}\times\hat{T}_{22}
^{\frac{1}{2}}\times\sin(\hat{T}_{11}^{\frac{1}{2}}\times\hat{T}_{22}^
{\frac{1}{2}}\times\theta_3),
\]
\begin{equation}
y' =
\hat{x}\times\hat{T}_{11}^{\frac{1}{2}}\times\hat{T}_{22}^{-\frac{1}{2}}
\times\sim(\hat{T}_{11}^{\frac{1}{2}}\times\hat{T}_{22}^{\frac{1}{2}}\times
\theta_3) +
\hat{y}\cos(\hat{T}_{11}^{\frac{1}{2}}\times\hat{T}_{22}^{\frac{1}
{2}}\times\theta_3),
\label{eq:six-11}\end{equation}
(see~[3h] for general isorotations in all three Euler angles).

As  one  can verify, isotransforms (\ref{eq:six-11}) leave invariant all
infinitely   possible   ellipsoidical   deformations   of   the sphere
$x\times x+y\times y+z\times z=r$     in the Euclidean space
$E(r,\delta,R),
r={x, y, z}, \delta = diag.(1, 1, 1)$,

\begin{equation}
r^t\times\delta\times r = x\times  \hat{T}_{11}\times x +
y\times \hat{T}_{22}\times y +
z\times \hat{T}_{33}\times z = r.
\label{eq:six-12}\end{equation}
In the {\sl isoeuclidean spaces}

\begin{equation}
\hat{E}(\hat{r},\hat{\delta},\hat{R}), \hat{r} = \{\hat{r}^k\}, \;
\hat{\delta}
=          \hat{T}_s\times\delta,        \;         \hat{T}_s        =
diag.(\hat{T}_{11},\hat{T}_{22},
\hat{T}_{33}), \; \hat{I}_s = \hat{T}_s^{-1},
\label{eq:six-13}\end{equation}
ellipsoid~(\ref{eq:six-12})     become  perfect     spheres
$r^{\hat{2}}   =
(r^t\times\delta\hat  {\times}r)\times\hat{I}$ called {\sl
isospheres}~[4g].

In fact, the lifting   of the semi axes  $1_k\rightarrow\hat{T}_{kk}  $
while the  related  units  are lifted  of  the {\sl   inverse} amounts
$1_k\rightarrow\hat{T}_{kk}^{-1}$, preserves the perfect
sphericity. This isoshericity is the  geometric  origin  of  the
isomorphism $\hat{O}(3)
\approx O(3)$, as well as
of  the preservation of the {\sl   rotational} invariance for the {\sl
ellipsoidical deformations} of sphere~[4b].

The connected {\sl isolorentz symmetry}  $S\hat{O}(3.1) $ (also called
in the literature {\sl Santilli's iso\-lo\-rentz symmetry}~[6,7], is
given
by the superposition of
the isorotations   and  the  {\sl    isoboosts} first  introduced
in~[4a]
which can be written in the (3,4)-plane

\begin{eqnarray}
x^1\prime &=&   x^1,\;\;\;   x^2\prime      =   x^2,  \nonumber  \\
x^3\prime &=&  
x^3\times\sinh(\hat{T}_{33}^{\frac{1}{2}}\times\hat{T}_{44}^
{\frac{1}{2}}\times v) -
x^4\times\hat{T}_{33}^{-\frac{1}{2}}\times\hat{T}_
{44}^{\frac{1}{2}}\times\cosh(\hat{T}_{33}^{\frac{1}{2}}\times\hat{T}_{44}^
{\frac{1}{2}}\times v) =  \nonumber      \\
&=&\hat{\gamma}\times
(x^3-\hat{T}_{33}^{-\frac{1}{2}}\times\hat{T}_{44}^
{\frac{1}{2}}\times\hat{\beta}\times x^4), \nonumber    \\
x^4\prime &=& -x^3\times\hat{T}_{33}^{\frac{1}{2}}\times c_0^{-1}\times
\hat{T}_{44}^{-\frac{1}{2}}\times\sinh(\hat{T}_{33}^{\frac{1}{2}}\times\hat{T}_{44}^{\frac{1}{2}}\times
v) +x^4\times\cosh(\hat{T}_{33}^{\frac{1}{2}}\times
\hat{T}_{44}^{\frac{1}{2}}\times v) = \nonumber \\
&=&\hat{\gamma}\times(x^4-\hat{T}_{33}^{\frac{1}{2}}\times\hat{T}_{44}^{-
\frac{1}{2}}\times\hat{\beta}\times x^3,
\label{eq:six-14}\end{eqnarray}
where

\begin{equation}
\hat{\beta} = (v_k\times\hat{T}_{kk}\times
v_k/c_0\times\hat{T}_{44}\times c_0)^
{\frac{1}{2}},
\label{eq:six-15a}\end{equation}
\begin{equation}
\hat{\gamma} = (1-\hat{\beta}^2)^{-\frac{1}{2}}.
\label{eq:six-15b}\end{equation}

Note that the above isotransforms  are {\sl nonlinear in} $x, \dot{x},
\psi,
\partial\psi, ..., $
precisely  as  desired,  and  are  formally  similar  to  the  Lorentz
transforms,  as  expected from their    isotopic character. This  also
confirms  the  local  isomorphism  $S\hat{O}(3.1)  \approx   SO(3.1) $
[4].

The Lorentz--Santilli isotransforms   characterize  the so--called  {\sl
isolight cone}~[4], i.e., the perfect cone in isospace $\hat{M}(\hat{x},
 \hat{\eta}, \hat{R})$. In a
way similar to the  isosphere, we have  the  deformation of the  light
cone axes $1_\mu\rightarrow \hat{T}_{\mu\mu} $
while the  corresponding units are  deformed of the {\sl
inverse}   amount $ 1_\mu \rightarrow \hat{T}_{\mu\mu}^{-1}$,
thus  preserving  the   original perfect cone
character. Such a preservation is then the  geometric foundation of the
local isomorphism $S\hat{O}(3.1) \approx SO(3.1)$.

The  isotopy  of   the  light cone  is   so   strong   that  even  the
characteristic angle of  the cone remains  the conventional one, i.e.,
{\sl the maximal  causal speed in isospace   $\hat{M}(\hat{x},
\hat{\eta},
\hat{R})$ remains  the speed of
light $c_0 $ in vacuum}~[4]
(it should be noted that the proof of this property requires, for
consistency,   the  use  of  the   isotrigonometric  and isohyperbolic
functions we cannot review here for brevity).

The {\sl isotranslations} in the coordinates can be written~[4d]

\begin{equation}
x' = (\hat{e}^{i\times p\times a})\hat{\times} x = x +
 a\times A(x, ...), \;\;\;
p' = (\hat{e}^{ip\times a})\hat{\times} p = p,
\label{eq:six-16a}\end{equation}
\begin{equation}
A_\mu = \hat{T}_{\mu\mu}^{1/2} +
a^\alpha[\hat{T}_{\mu\mu}^{1/2},\hat{}p_
\alpha]1! + ...
\label{eq:six-16b}\end{equation}

It  is generally  believed   that  the conventional, ten-parameter,
Poincar\'{e}  symmetry  is  the
broadest possible linear  symmetry of  the conventional separation  on
Minkowski spaces $M(x, \eta, R) $

\begin{equation}
(x - y)^2 = [(x^\mu -
y^\mu)\times\eta_{\mu\nu}\times(x^\nu-y^\nu)]\times I
\in R(n,+,\times).
\label{eq:six-17}\end{equation}

The  isotopies have identified a new
symmetry,  called
{\sl isoselfscalarity}, first identified in memoir [3g], which
which is given  by the lifting  of the trivial unit
$I = diag.(1, 1, 1, 1)$ with a new parameter $n$ independent from the
integration
variables, under which we have the new invariance

\[
I \rightarrow \hat{I} = n^2\times I, \;\;\; \eta \rightarrow \hat{\eta}
=
n^{-2}\times\eta, \; n \not = 0,
\]
\begin{eqnarray}
\lefteqn{(x - y)^2 = [(x^\mu -
y^\mu)\times\eta_{\mu\nu}\times(x^\nu-y^\nu)]
\times I \equiv } \nonumber \\
& &\equiv
[(x^\mu-y^\mu)\times(n^{-2}\times\eta_{\mu\nu})\times(x^\nu-y^\nu)]
\times(n^2\times I) \nonumber \\
& & = [(x^\mu-y^\mu)\times\hat{\eta}_{\mu\nu}
\times(x^\nu-y^\nu)]\times \hat {I} =
(x-y)^{\hat{2}}.
\label{eq:six-18}\end{eqnarray}

As a result, the most general possible invariance of the Minkowskian
line
element for positive-definite units has
{\it eleven}, rather than {\it ten} dimensions.

Note that the  invariant for the first form
of the line element is the conventional  Poincar\'{e} symmetry, while
the
invariance of  the latter  form  is a  bona--fide isopoincar\'{e}
symmetry
because the  isotopic element $\hat{T} = n^2 $  enters into the {\sl
arguments} of
the isorotation~(\ref{eq:six-11})  and   isoboosts~(\ref{eq:six-14}). As
such,    the above
symmetry is nontrivial.

A second, hitherto unknown invariance is characterized by the {\sl
isodual map} [5d]

\[
I \;\rightarrow\; I^d = -I,\; \eta \;\rightarrow\; \eta^d = -\eta,\;
x\;\rightarrow\; x^d = -x,
\]
\begin{eqnarray}
\lefteqn{(x - y)^2 = [(x^\mu -
y^\mu)\times\eta_{\mu\nu}\times(x^\nu-y^\nu)]
\times I \equiv } \nonumber \\
& & \equiv
[(x^\mu-y^\mu)\times(-\eta_{\mu\nu})\times(x^\nu-y^\nu)]\times(-I) =
 \nonumber \\
& & =[(x^\mu-y^\mu)^d\times\eta_{\mu\nu}^d\times(x^\nu-y^\nu)^d]\times
I^d
\equiv (x-y)^{d2d}.
\label{eq:six-19}\end{eqnarray}
called by this author {\sl isoselfduality}, which is at the foundation
of the {\sl isodual representation of  antimatter}~[5].

The above invariance evidently assures the plausibility of the isodual
treatment of antimatter also at the geometric  level, because the same
Minkowskian invariant holds for both conventional and isodual systems.

Isoselfduality~(\ref{eq:six-19})  establishes the existence and
applicability of
the  {\sl isodual Poincar\'{e}  symmetry} $\hat{P}^d(3.1)$ [4d],
which  can be  easily constructed
from the isodual rules  of  the preceding section, and it is  hereon
assumed as known.

We  here define as the   {\sl  restricted isopoincar\'{e} transforms}
that
constituted by  isorotations, isolorentz boosts,  isotranslations,
isoinversions and isoselfscalar transformations when all
parameters are constants, otherwise we have the
{\sl    general isopoincar\'{e}
tramnsforms}, with corresponding definitions under isodualities.

The most salient difference between the special and general
isotransforms
is that the former preserve the inertial charactyer of the frames, while
the latter identify a broader class of noninertial frames.

Note that isodual parameters are independent of the conventional ones.
As a result, the general invariance of isoseparation~(\ref{eq:three-5b})
has 22--dimensions with structure

\begin{equation}
\hat{S}_{\rm dot} =
\hat{P}(3.1)_{|\hat{I}}\times\hat{P}^d(3.1)_{|\hat{I}^d}
\label{eq:six-20}\end{equation}
and the same result holds for the symmetry of the conventional
Minkowskian
separation as a particular case.

The preceding analysis  establishes the following property [4,7a]

\vskip 1cm
{\bf Theorem 2:} {\sl The  general isopoincar\'{e} group is the
universal
isolinear  invariance  of  all  infinitely   possible,  well   behaved
isoseparations~(\ref{eq:two-5b}) on  the  isominkowski space over
isoreal  fields
with a common isounit}.
\vskip 1cm

The verification of the  above theorem is trivial and  can be done  by
just plotting the isotransforms in isoinvariant~(\ref{eq:three-5b}).
Note that {\sl
there is nothing to compute}, because the theory provides the solution
for the general   invariance of the  isoseparation  for all infinitely
possible isometrics  in the diagonal form $\hat{\eta} =
\hat{T}\times\eta =
diag.(T_{\mu\mu}\times\eta_{\mu\mu})$. One {\sl merely plots}
the isotopic elements $T_{\mu\mu}$ in  the isotransforms. The invariance
of the
isoseparation is then guaranteed by the isotopic methods.

Note the need for  consistency that {\sl  the generalized unit must be
the same for the isospace and for the  isofield}. This is not the case
in  conventional treatments  where  the unit  of the space  is the unit
matrix  $I = Diag. (1, 1, ...)$   and the unit  of  the  field is  the
number
+1. Nevertheless, the latter treatment  can be easily reformulated for
the same unit I.

Note  also from the  above studies that  {\sl the  abstract identity of
the
Poincar\'{e} and  isopoincar\'{e} symmetries  implies that  the  special
and
isospecial   relativity also coincide  at the  abstract level (and the
same occurrence  holds    under  isoduality)}. However,   the  special
relativity has only  {\sl  one formulation}, the conventional  one. On
the contrary, all isotopic structures,  thus including the  isospecial
relativity, have {\sl two different formulations}, one in isospace and
the other in their projection in the conventional space.

All differences between the special  and isospecial relativity  solely
occur when  the latter is projected   in the space--time of  the former,
because the isospecial relativity in  isospace preserves all  features
of the conventional relativity,  including the perfect light cone, the
maximal  causal speed $c_0$, etc.~[4].

It should be also noted that
that {\sl  the  general isopoincar\'{e}  symmetry does not
restrict  the value of} n (except  for the conditions  $n>0$). Thus,
{\sl
the isospecial relativity   predicts arbitrary causal speeds  of light
within homogeneous   and isotropic media},  because c  can be smaller,
equal or bigger than $c_0 $.

The   case  of light  speeds   smaller  than  $c_0  $  is established
in
homogeneous and  isotopic physical media  such as water. Speeds bigger
than $c_0$  have been identified  in a number  of cases, such as photons
tunneling  tests~[15a,15d], expulsion   of matter  in  astrophysical
explosions~[15c,15d,15e],  solutions   of ordinary   relativistic   wave
equations~[15f] and other cases.

The isominkowskian space, the isopoincar\'{e} symmetry and the
isospecial
relativity restore the validity of  the abstract axioms of the special
relativity for  arbitrary speeds of light,  whether  smaller or bigger
than $ c_0$ ,because they are all projected  in isospace into the unique
and universal speed $c_0$, thus rendering  the special relativity truly
"universal".

It  is remarkable  that the two   novel invariances~(\ref{eq:five-24})--
~(\ref{eq:five-25}) and
(\ref{eq:six-18})--(\ref{eq:six-19}) have remained
undetected throughout this century, to our
best knowledge.  This should not be surprising because their detection
required the prior discovery of {\sl new  numbers, the isonumbers with
arbitrary positive units} for  invariances~(\ref{eq:five-24})
and~(\ref{eq:six-18}) and  the
{\sl  isodual numbers with arbitrary  negative  units} for
invariances~(\ref{eq:five-25}) and (\ref{eq:six-19}).

\section{ Operator isominkowskian gravity for matter and its isodual for
antimatter}
\setcounter{equation}{0}

We  are now sufficiently  equipped to identify  the foundations of OIG
and appraise its plausibility on comparative grounds with QG [12]. A
number
of   applications and verifications  are  presented   in future
works~[4i].
OIG was first presented in ref.~[8a]  and preliminary studied
in~[8b,8c].
The  basic   dynamical   equations of   OIG   in their
axiomatically correct  form are presented in  this paper for the first
time following the achievement of mathematical maturity in memoir
[3f] and axiomatic maturity in memoir [3g].

We should indicate from the outset that the
expectation  of the existence of a  "quantum"  description of gravity
is disproved by ouur studies.  This  is established  by Theorem   1
which
identifies the classical   {\it noncanonical}  structure of  gravity
with a
necessary {\it nonunitary} counterpart at the operator level, under
which no
"quantum" law is expected to apply identically.

In  fact,  QG  requires  notorious departures   from   a true "quantum
mechanical" setting and the isotopic reformulation of gravity does not
escape from the  same occurrence. The best  that can be done on ground
of our current  knowledge is a formulation   of gravity via  the {\it
abstract axioms of quantum mechanics,  only realized in a more general
form}.

Equivalently we can say that the main idea of OIG is to turn the
notorious "nonunitary" structure of the operator description
of gravity on a conventional Hilbert space, into an identical
"isounitary"
formulation on our isohilbert space, thus regaining in this way
the abstract axioms of RQM.

Alternatively, our studies  indicate  that no formulation  of  gravity
appears to be possible via the use of the quantum axioms {\sl in their
simplest possible realization}, that with the unit +1. On the contrary,
if more general {\sl realizations} are admitted, then
realistic possibilities for basic advances emerge.

>From  the analysis of  the preceding sections it  is evident that RHM,
with underlying  isominkowskian geometry, isopoincar\'{e}  symmetry
and  isospecial relativity,  provides an operator  characterization of
gravity    under the sole   condition  of restricting  the isounit and
isotopic element to the gravitational values

\begin{equation}
\hat{T}_{gr}(x)_{\mu^\nu} = \eta_{\mu\alpha}\times g^{\alpha\nu}(x),
\label{eq:seven-1a}\end{equation}
\begin{equation}
\hat{I}_{gr}(x) = (\hat{I}_{gr}(x)^\mu_\nu) = (|\hat{T}_{gr}(x)_\alpha^
\beta|^{-1})^\mu_\nu, \;\; \eta \in M, \;\;\;, \;\; g \in \Re.
\label{eq:seven-1b}\end{equation}
called {\sl gravitational isotopic  element} and  {\sl gravitational
isounit, respectively}.

Conventional    RQM represents  systems  via   the assignment    of the
Hamiltonian H and the {\sl tacit  assumption} of the simplest possible
unit  I.  OIG requires  the  assignment  of {\sl two}  quantities, the
conventional Hamiltonian H which represents conventional interactions,
and the selection  of the isounit $\hat {I} = \hat{I}_{gr}(x) =
[\hat{T}_
{gr}(x)]^{-1} $  which  represents the essential
part of   curvature,   the isotopic  element  in  our   isominkowskian
decomposition of the  Riemannian metric $g(x) = \hat{T}_{gr}\times\eta$.
We assume the  reader is
now  familiar with the  mathematical  structure of OIG which  requires
{\sl all} products to be isotopic with isotopic element
$\hat{T}_{gr}(x)$.

Recall that the fundamental   notion of RQM  is the  Poincar\'{e}
symmety
P(3.1) in operator realization. By the same  token, the ultimate and
most  fundamental notion of the operator  theory herein submitted from
which   all   properties   and  applications  can   be    uniquely and
unambiguously  derived,  is the  {\sl gravitational
Poincar\'{e}--Santilli
isosymmetry in operator form} $\hat{P}_{gr}(3.1)$,
i.e., the isosymmetry of the
preceding   section constructed  with  respect  to   the gravitational
isounit $\hat{I}_{gr}(x)$ .

The main   characteristics of the  emerging  operator gravity  are the
following:

\vskip 1cm
{\bf Property I:} {\sl OIG is based on the embedding of gravity in the
isotopic lifting of Planck's constant}.
\vskip 1cm

Recall  that the  Plank constant   is  the  basic  unit  of RQM.   The
fundamental  isotopy of OIG is then  precisely that of the latter, and
we shall write

\begin{equation}
\hbar = I\; \rightarrow\;\hat{I}_{gr}(r),
\label{eq:seven-2}\end{equation}
where  $\hat{I}_{gr}(r)$  is  the $3\times 3$--dimensional
{\sl  space   component} of the
$4\times 4$--dimensional gravitational isounit $\hat{I}_{gr}(x)$.

The isotopic character of  the lifting is  readily established  by the
fact that $\hat{I}_{gr}(r)$ preserves all axiomatic properties of
$\hbar$,
e.g.,

\begin{equation}
\hat{I}_{gr}^{\hat{n}} =
\hat{I}_{gr}\hat{\times}\hat{I}_{gr}\hat{\times}
..\hat{\times}\hat{I}_{gr}\equiv\hat{I}_{gr},\;\;\hat{I}_{gr}^{\hat{
\frac{1}{2}}}\equiv\hat{I}_{gr},
\;\; \hat{I}_{gr}\hat{/}\hat{I}_{gr}\equiv\hat{I}_{gr},
\label{eq:seven-3}\end{equation}

The fundamental  dynamical equations of  OIG are then based on
lifting~(\ref{eq:five-2}).    Note that   the {\sl conventional}
Schr\"{o}dinger  and
Heisenberg's equations can be written in the form

\begin{equation}
i \times \partial_t|\psi> = H(t,r,p)\times\hbar^{-1}\times|\psi>=
E\times\hbar^
{-1}\times|\psi>,
\label{eq:seven-4a}\end{equation}
\begin{equation}
i \times dA / dt = A \times \hbar^{-1} \times H - H \times \hbar^{-1}
\times A,
\label{eq:seven-4b}\end{equation}
\begin{equation}
p_k\times\hbar^{-1}\times|\psi> = -i\times \partial_k|\psi>,
\label{eq:seven-4c}\end{equation}
\begin{equation}
p_i\times\hbar^{-1}\times r^j - r^j\times\hbar^{-1}\times p_i = -
\delta_i^j
\label{eq:seven-4d}\end{equation}
\begin{equation}
p_i\times\hbar^{-1}\times p_j - p_i\times\hbar^{-1}\times p_i \equiv r^i
\times\hbar^{-1}\times r^j - r^j\times\hbar^{-1}\times r^i \equiv 0.
\label{eq:seven-4e}\end{equation}
Then, the {\sl fundamental non--isorelativistic dynamical equations} of
OIG are given by

\begin{equation}
i \times \hat{\partial}_t|\psi> =
\hat{H}(\hat{t},\hat{r},\hat{p})\times\hat{T}_{gr}(r)
\times|\hat{\psi}> = \hat{E}\times\hat{T}_{hr}(r)\times|\hat{\psi}>,
\label{eq:seven-5a}\end{equation}
\begin{equation}
i\times \hat{d}\hat{A} / \hat{d}\hat{t} =
\hat{A}\times\hat{T}_{gr}(r)\times\hat{H} -
\hat{H}\times\hat{T}_{gr}(r)\times\hat{A},
\label{eq:seven-5b}\end{equation}
\begin{equation}
p_k\times\hat{T}_{gr}(r)\times|\hat{\psi}> = -i\times
\hat{\partial}_k|\hat{\psi}>,
\label{eq:seven-5c}\end{equation}
\begin{equation}
\hat{p}_i\times\hat{T}_{gr}(r)\times r^j -
r^j\times\hat{T}_{gr}(r)\times p_i =
-i\times \delta_i^j
\label{eq:seven-5d}\end{equation}
\begin{equation}
p_i\times\hat{T}_{gr}(r)\times p_j - p_j\times\hat{T}_{gr}(r)\times
p_i\equiv
r^i\times\hat{T}_{gr}(r)\times r^j - r^j\times\hat{T}_{gr}(r)\times r^i
\equiv 0,
\label{eq:seven-5e}\end{equation}
where the  isounit of the time isoderivatives  is evidently the fourth
component of $\hat{I}_{gr}(x)$, under the    proviso that the  {\sl
totality}  of
quantities, operations and special
functions and transforms are of isotopic type.

To be more specific on this fundamental point, the appraisal of OIG with
conventional QG notions, such as the magnitude of the angular momentum
$J^2 = J_k\times J^k$ leads to a host of inconsistencies which are
generally
not detected by nonexpert in the field (e.g., violation of
isolinearity).
Similarly, data elaborations via ordinary trigonometric functions or
with the familiar Dirac's delta function have no meaning of any nature
for
OIG, because said conventional notions cannot be even defined in
isospaces over isofields.

With a clear understanding to above requirements, we note that
{\sl the gravitational isounit is inneed the fundamental invariant of
OIG}
because it is {\sl numerically invariant} under the transformation
theory

\begin{equation}
\hat{I}_{gr} \rightarrow \hat{I}_{gr}' =
\hat{U}\hat{\times}\hat{I}_{gr}\hat
{\times}\hat{U}^\dagger \equiv \hat{I}_{gr},
\label{eq:seven-6}\end{equation}
and it is preserved under the time evolution

\begin{equation}
i\times \hat{dI}_{gr} / dt = [\hat{I}_{gr}, \hat{}\hat{H}] = \hat{H} -
\hat{H} \equiv 0.
\label{eq:seven-7}\end{equation}

Moreover, from isorule~(\ref{eq:three-10}), the isoexpectation values of
the space
components of the gravitational isounit reproduce Plank's constant
$\hbar = 1$ identically,

\begin{equation}
<\hat{I}_{gr} = \frac{<|\times\hat{T}_{gr}\times\hat{T}_{gr}^{-1}\times
\hat{T}_{gr}\times|>}\times\hat{T}_{gr}\times|> \equiv \hbar = 1. =
\hbar = 1.
\label{eq:seven-8}\end{equation}

This identifies the ``hidden'' character of OIG in conventional RQM and
its character of being a ``completion'' of RQM much along the E--P--R
argument~[14]. After all, {\sl gravity is embedded in the unit of RQM}.
As such, the inclusion of gravity in RQM is so natural to creep in
un--noticed.

Also, property~(\ref{eq:seven-8}) establishes that one {\sl should not }
expect OIG
to yield  {\sl  deviations} from  established  quantum mechanical
laws.  This
occurrence is made clearer by the fact  that {\sl the uncertainties of
the center--of--mass trajectories of a  systems of particles obeying OIG
are conventional}. In fact, from isocommutation rules~(\ref{eq:five-5d})
we have $(\hbar =
1)$ [3g]

\begin{equation}
\Delta r\times\Delta p \geq \frac{1}{2} <[\hat {r},\hat{}p]> =
\frac{1}{2}.
\label{eq:seven-9}\end{equation}

The  {\sl preservation} of the fundamental physical laws by our
"axioms-- preserving" {\sl isotopies} should  be compared with the {\it
departures} from    the same laws implied by QG as well as  by the
"axiom--violating" {\sl   deformations},  and   illustrates  again  our
insistence  in   avoiding the term "deformations"
whenever dealing with "isotopies" (Sect. 2).

 The  presence  of gravitational IN OUR OIG is established by
numerous aspects, all verifying conventional quantum laws, such as
deviations
from conventional quantum eigenvalues, or
the resolution of {\sl the paradox of
quantum mechanics at gravitational singularities}~[8d] (see below).

The {\sl isorelativistic equations}  of  OIG are uniquely
identified by the  iso\-po\-in\-ca\-r\'{e} symmetry via its isocasimir
invariants~(\ref{eq:six-10})
and related isorepresentation  theory we cannot  possibly study
here for brevity (see the study of ref.~[44h,7a]). The first difference
is that now the gravitational
isounit is given by the full $4\times 4$--dimensional structure
$\hat{I}_{gr}
(x)$. The {\it fundamental gravitational isocommutation rules} are given
by

\begin{equation}
p_\mu\hat{\times}|\hat{\psi}> =
p_\mu\times\hat{T}_{gr}\times|\hat{\psi}> =
-i\times\hat{\partial}_\mu|\hat{\psi}> =
-i\hat{T}_{{gr}\mu}^{\nu}\partial_\nu|\hat{\psi}>,
\label{eq:seven-10a}\end{equation}
\begin{equation}
[x^\mu,\hat{}p_\nu]\hat{\times}|\hat{\psi}> =
[\hat{x}^\mu\times\hat{T}_{gr}(x)
\times\hat{p}_\nu - \hat{p}_\nu\times\hat{T}_{gr}(x)
\times\hat{x}^\mu]\times\hat{T}_{gr}(x)\times|\hat{\psi}> =
i\times \delta^\mu_\nu\times|\hat{\psi}>,
\label{eq:seven-10c}\end{equation}

The {\sl second--order isorelativistic equation of QIG} is then given by
the
realization of the isoinvariant~(\ref{eq:four-10b}) plus the
conventional minimal coupling
rule to an external electromagnetic field with four-potential
$\hat{A}_\mu(x)$

\begin{equation}
\!\!\!\!\!\!\!\!\{[\hat{p}_\mu + i\times
e\times\hat{A}_\mu(x)]\hat{\times}[\hat{p}^\mu +
i\times \times e\times\hat{A}^\mu(x)]
+\hat{m}^{\hat{2}}\}\hat{\times}|\hat{\psi}>=
\end{equation}
\begin{equation}
\eta_{gr}^{\hat{}\mu\nu}(x)\times[\hat{p}_\mu + i\times
e\times\hat{A}_\mu(x)]
\times[\hat{p}_\mu + i\times \times e\times\hat{A}_\nu(x)]+
\end{equation}
\begin{equation}
+ (m\times m)\times\hat{I}_
{gr}(x)\}\times\hat{T}_{gr}(x)\times|\hat{\psi}> =
\end{equation}
\begin{equation}
=\{\hat{I}_{gr}(x)^\mu_\alpha\times \eta^{\alpha\nu}[-i\times
\hat{T}_{gr}(x)_\mu^\gamma
\times\partial_\gamma+i\times e\times\hat{A}_\mu(x)]\times
\end{equation}
\begin{equation}
\times [-i\times \hat{T}_{gr}(x)_\nu^
\delta\times\partial_\delta+ie\times\hat{A}_\nu(x)]+m^2\}\times|\hat{\psi}>=
\end{equation}
\begin{equation}
=\{\eta^{\rho\sigma}\times [-i\times \partial_\rho+i\times
e\times\hat{I}_{gr}(x)^\mu_\rho
\times\hat{A}_\mu(x)]\times
\end{equation}
\begin{equation}
\times [-i\times\partial_\sigma+ie\hat{I}_{gr}(x)^\nu_\sigma
\times\hat{A}_\sigma(x)]+m^2\}\times|\hat{\psi}> = 0,
\label{eq:seven-11}
\end{equation}

where quantities with "hats" are computed in isospace and those without
are
conventional.

As one can see, the projection of dynamical equation~(\ref{eq:seven-11})
in the conventional
space--time can be expressed via the conventional Klein--Gordon equation
{\sl plus} the "isorenormalization" of the electromagnetic potential $
\hat{A}_{\mu}$ via
the multiplication by the gravitational isounit $\hat{I}_{gr}(x)$.
Conceivable physical
applications of this new setting will be studied elsewhere.

The "isolinearization" of the second--order equation is done via a
simple isotopy
of the conventional case (see~[4h] for details). In its simplest
possible
realization for a diagonal isounit, such linearization leads to the
following
{\sl gravitational isodirac equation}

\begin{equation}
(\hat{\gamma}^\mu\hat{\times}\hat{p}_\mu + i\times
\hat{m})\hat{\times}|> =
[\eta_{\mu\nu}(x)\times\hat{\gamma}^\mu(x)\times\hat{T}_{gr}(x)\times\hat{p}^\nu
-
i\times m\times\hat{I}_{gr}(x)]\times\hat{T}_{gr}(x)\times|> = 0,
\label{eq:seven-12a}\end{equation}
\begin{equation}
\{\hat{\gamma}^\mu, \hat{}\hat{\gamma}^\mu\} =
\hat{\gamma}^\mu\times\hat{T}_{gr}(x)\times\hat{\gamma}^\nu +
\hat{\gamma}^\nu\times\hat{T}_{gr}(x)\times\hat{\gamma}^\mu =
2\times\hat{\eta}^{\mu\nu}\equiv 2 g^{\mu\nu},
\label{eq:seven-12b}\end{equation}
\begin{equation}
\hat{\gamma}^k =
[\hat{T}_{kk}(x)]^{1/2}\times\gamma^k\times\hat{I}_{gr}(x) =
[\hat{T}_{kk}(x)]^{1/2}\times
\left ( \begin{array}{cc}
0 & \sigma^k \\
\sigma^{d k} & 0 \end{array} \right ) \times\hat{I}_{gr}(x)
\label{eq:seven-12c}\end{equation}
\begin{equation}
\hat{\gamma}^4 = [\hat{T}_{44}(x)]^{1/2}\times\gamma^4\times
\hat{I}_{gr}(x) =
[\hat{T}_{kk}(x)]^{1/2}\times \left( \begin{array}{cc}
I_{2\times 2} & 0 \\
0 & I_{2\times 2}^d \end{array} \right )\times\hat{I}_{gr}(x)
\label{eq:seven-12d}\end{equation}
with a simple extension with the minimal coupling rule, where the
$\gamma$'s
 are the
conventional Dirac matrices, the $\hat {\gamma}$'s are the {\sl isodirac
matrices}, and
the symbol d stands for {\sl isoduality}, i.e., $\sigma^d =
-\sigma^\dagger =
-\sigma,\: I^d = -I$.

As one can {\sl see, the anti--isocommutators of the isogamma matrices
yield
(twice) the Riemannian metric} $g(x)$, thus confirming the
representation of
gravitation in the {\sl structure} of Dirac's equation with the
{\sl conventional} Riemannian metric $g(x)$, as desired. As an example,
we have
the particular case for the {\sl iso--Dirac--Schwarzchild equation}~[2d]

\begin{equation}
\hat{\gamma}_k = (1-2M/r)^{-1/2}\times\gamma_k\times\hat{I}_{gr}(x),
\;\;\;
\hat{\gamma}_4 = (1-2M/r)^{1/2}\times\gamma_4\times\hat{I}_{gr}(x).
\label{eq:seven-13}\end{equation}
Similar isorelativistic gravitational equations can be easily
constructed by
the interested reader.

It  is generally believed  that the conventional Poincar\'{e} symmetry
in its spinorial  covering ${\cal P}(3.1)  = SL(2.c)\times  T(3.1)$ is
the general  symmetry of the conventional  Dirac equation. This belief
can be disproved by the isodual mathematics~[5]. In fact, we have the
following

\vskip  1cm  {\bf  Theorem  3:} {\sl  The   largest possible isolinear
symmetry   of the  isodirac equation is     given by the  isospinorial
isopoincar\'{e} symmetry in the following 22-dimensional isoselfdual
form}

\begin{equation}
\hat{S}_{tot} = \hat{\cal P}(3.1)\times\hat{\cal P}^d(3.1) =
[S\hat{L}(2,\hat{C})\hat {\times}\hat{T}(3.1)]\hat
{\times}[S\hat{L}^d(2.\hat{C}^d)
\hat {\times}\hat{T}^d(3.1)],
\label{eq:seven-14}\end{equation}

\vskip 1cm
{\bf Prof.} The conventional gamma matrices are isoselfdual, i.e.,
invariant
under isoduality, $\gamma_\mu\equiv\gamma_\mu^d = -\gamma_\mu^\dagger$.
A necessary condition for a Lie transformation group to
be a symmetry of the conventional Dirac's equation is therefore that it
must
also be isoselfdual for consistency. The isospinorial Poincar\'{e}
symmetry
${\cal P}(3.1)$ is not isoselfdual, ${\cal P}^d(3.1)\not = {\cal
P}(3.1)$,
and therefore does not verify the indicated
necessary condition. However the direct product $S_{tot} = {\cal
P}(3.1)\hat {\times}
{\cal P}^d(3.1)$ is isoselfdual, $S_{tot}\equiv S_{tot}^d$, and
therefore verifies this necessary condition. The sufficiency can be
proved as
in the conventional case.
Since the isotopies are axiom--preserving,
the above properties of the {\sl conventional} Dirac equation persists
under
all its infinitely possible isotopies, including the gravitational
particularization. Finally, the 22-dimensional character of the total
symmetry
originates from the independence of conventional and isodual parameters,
as well
as the inclusion of the isoselfscalar transforms and their isoduals of
Theorem 2 of Sect. 6 {\bf q.e.d.}

Theorem 3 can also be reached by an inspection of the {\sl conventional}
Dirac
matrices in Eq.s~(\ref{eq:seven-14b}) and (\ref{eq:seven-14c}).
In fact the latter are centrally dependent
on {\sl the negative $2\times 2$--dimensional unit
for the internal space of spin} which
 we have merely rewritten in our isodual form $I_{2\times 2}^d =
-I_{2\times
2} = -Diag. (1, 1)$.
This illustrates that the birth of the isodual theories for
antiparticles can
be seen in the {\sl conventional} Dirac's equation because of the
essential
presence of a negative unit in the very structure of the gamma matrices.

The latter was not interpreted by Dirac as a {\sl bona--fide} unit
because he
lacked the knowledge of the related {\sl new numbers with negative
units, the
isodual numbers}~[3e].

Similarly, the space part of the {\sl conventional} Dirac matrices
reveals the
presence {\sl of the isodual Pauli matrices, only written in our
formalism}
$\sigma^{kd} = -\sigma_k^\dagger = -\sigma_k$.
The latter occurrence has intriguing implications, such as:

a) It establishes the validity of the isodual representation of
antimatter.
In fact, the referral of the negative--energy solutions to negative
units
eliminates their un--physical behavior~[5];

b) It establishes the insufficient character of current interpretation
of
${\cal P}(3.1)$
as being the maximal linear symmetry of the Dirac equations, in favor of
the isoselfdual form ${\cal P}(3.1)
\times {\cal P}^d(3.1)$.

c) It disproves another popular belief according to which {\sl th spin
in  Dirac  matrices is  characterized by a $4\times  4$--dimensional
representation} of SU(2). In  fact, the treatment   of spin is restored
as  being
entirely characterized  by  the 2--dimensional  Pauli's  representation
{\sl for the case  of particles}, with the independent antiautomorphic
isodual   Pauli's matrices for  the characterization   of the spin  of
antiparticles,  again, along the   isoselfdual structure $SU(2) \times
SU^d(2)  \in  {\cal  P}(3.1) \times{\cal   P}^d(3.1)$. At any rate,
SU(2) is not
isoselfdual and, as such, it cannot consistently characterize the spin
of
Dirac's equation. For  additional
studies along these  lines we  refer the  interested reader  to
ref.s~[4h,5].

\vskip 1cm {\bf Property II:} {\sl OIG  coincides at the abstract level
with RQM.}

\vskip 1cm In view of  the positive--definiteness of the gravitational
isounit  $\hat{I}_  {gr}(x)$  (originating from  the local Minkowskian
character of the (3+1)--dimensional Riemannian spaces, Sect. 1), at the
abstract level  we have  the  identity of  $I$ and  $\hat{I}_{gr}(x)$,
$\cal    H$    and     $\hat{\cal   H     },     R(n,+,\times)$    and
$\hat{R}(\hat{n},+,\hat{\times})$, etc.   The    same holds for    the
dynamical equations, e.g., at the abstract  level we have
$(\gamma^\mu\times
p_\mu  + i\times m)|>   \equiv  (\hat{\gamma}^\mu\hat{\times}\hat{p}_\mu
+
\hat {i}\hat{m})\hat{\times}|>$.

Needless to say,  the  above abstract  identity of  OIG and  RQM  {\sl
guarantees}  the axiomatic  consistency of OIG   and, and such, it is
simply invaluable for the resolution of the  problematic aspects of QG
(see below).

An inspection  of the  properties   of the  isopoincar\'{e}  symmetry,
particularly  the isocommutativity of   coordinates and  momenta,
Eq.s~(\ref{eq:four-5e}), established the following

\vskip 1cm {\bf Property III} {\sl OIG is isoflat (i.e., it verifies the
axiom of flatness in isospace).}

\vskip  1cm   By comparison,  QG   is  curved in    the sense that its
coordinates and momenta do not  commute. This difference is basic  for
the  understanding of  the differences of  the  two theories and their
consequential comparative appraisals.

In essence, we can state that {\sl the isocommutativity of coordinates
and momenta is a necessary condition for the unambiguous applicability
of the theory to measurements} (in view of Theorem 1).

The reader should  recall the {\sl  dual} formulation of  all isotopic
theories,   that  in   isospaces    and  its  projection in   ordinary
spaces. Therefore, the  isocommutativity of  coordinates and  momentum
does not imply that the theory is {\sl ordinarily flat}, in which case
no  representation of gravitation  is evidently  possible owing to the
historical teaching of  the Founders of  the theory~[2]. In fact,  the
projection of OIG in ordinary spaces over conventional fields recovers
all conventional Riemannian characteristics.

The distinction between {\sl isoflatness}  and {\sl ordinary flatness}
is here merely indicated with technical treatments presented elsewhere
for brevity~[9].

An important  implication of  the   above  studies is  that {\sl   the
isospecial  relativity can   indeed  unify  the special  and   general
relativities in both their    classical and quantum versions  for  the
exterior   gravitational  problem of matter   in  vacuum, with isodual
images for  the exterior gravitational  problem  of antimatter}. The
relativities are unified for both  classical and operator versions via
the   basic unit which  can  represent  gravitation when  assuming the
gravitational form $\hat{I}_{gr}(x)$ and admits as particular case the
special relativity when assuming the trivial form $I = Diag. (1, 1, 1,
1)$, with isodual images for antimatter.

It should be noted that {\sl  the isospecial relativity can also unify
the  special and   general  relativities  for  interior  gravitational
problems of matter within physical  media in both their classical  and
operator versions,  with   isodual images for antimatter}.   The latter
unification  is permitted  by  the  unrestricted
functional dependence   of the  isounit which can   represent interior
gravitational conditions via realizations of the type

\begin{equation}
\hat{I}_{gr, int} = K(x, \dot{x}, \ddot{x}, ...)\times\hat{I}_{gr}(x),
\label{eq:seven-13a}\end{equation}
where K is a positive--definite $4\times 4$
 matrix representing arbitrary nonlinear and
nonlocal internal effects, the interior relativistic conditions emerging
for
the simpler value

\begin{equation}
\hat{I}_{rel, int} = K(x, \dot{x}, \ddot{x}, ...)\times I.
\label{eq:seven-15}\end{equation}

The simplest possible realization of the internal gravitational isounit
is
given by~[3g]

\begin{equation}
\hat{I} = [Diag. (n_1^{-2}, n_2^{-2}, n_3^{-2}, n_4^{-2})]\times F(x,
\dot{x},
 ...)\times\hat{I}_{gr}(x).
\label{eq:seven-16}\end{equation}
All conventional exterior gravitational models (e.g., Schwarzschild's
exterior
metric~[2]) can the be easily lifted to the above interior conditions.

The main physical result is the {\sl extension of general relativity to
locally varying speeds of light under the preservation of its abstract
axioms}.
In fact, the speed of electromagnetic waves in interior conditions is a
rather
complex function of local variables, such as
density $\mu$, temperature $\tau$, frequency $\omega$,
etc., $c = c(x, \mu, \tau, \omega, ...) = c_0/n(x, \mu, \tau, \omega,
..)$.
The important point is that the above local speed can be {\sl directly
represented}, that is, represented via the {\sl geometric line element
of the theory}, rather than current indirect manipulations (e.g., the
representation of the {\sl classical} speed $c = c_0/n$ via the
scattering of photons
among molecules in {\sl second} quantization).

As an example, it is well known that the representation of the locally
varying
character of the speed of light in interior conditions is not possible
in the
Schwarzschild's geometry. By comparison, its representation becomes
elementary
under our isotopic extension to interior conditions, due to the lifting
of the
fourth component (here assuming $F=I$ for simplicity)

\begin{equation}
g_{44} \rightarrow \hat{g}_{44} = g_{44}/n_4^2.
\label{eq:seven-17}\end{equation}
The space components  $n_k^2$   then emerge  as the {\sl    space--time
symmetrization of the index of refraction},  or via simple application
of the Lorentz--Santilli isoboosts.

The isominkowskian reformulation  of gravity also  permits a novel and
physically more accurate representation  of gravitational horizons and
singularities via the {\sl  zeros of the  isotopic element or isounit,
according to the rules}~[4h]

\[
{\rm Gravit.-Horizons:} \;\;\left\{ \begin{array}{l}
{\rm  Space-component-of}- \hat{I}_{gr,int} = Diag.
(\hat{I}_{11}, \hat{I}_{22}, \hat{I}_{33}) = 0, \\
{\rm Time-component-of}-\hat{T}_{gr,int} = \hat{T}_{44} = 0,
\end{array} \right.
\]
\begin{equation}
{\rm Gravitat.-Singularities:} \;\; \left\{ \begin{array}{l}
{\rm Time-component-of}- \hat{I}_{gr,int} = \hat{I}_{44} = 0. \\
{\rm Space-component-of}- \hat{T}_{gr,int} = Diag.(\hat{t}_{11},
\hat{T}_{22},
\hat{T}_{33}) = 0. \end{array} \right.
\label{eq:seven-18}\end{equation}

To understand there rules, note first that they are verified by the
Schwarzschild
 metric for which we have~[4h]

\begin{equation} \begin{array}{l}
{\rm Gravitational--Horizons:\;\; space--component--of--}\; g = (1-2 M/
r)
Diag.(1,1,1) = 0,\\
{\rm Gravitational--Singularity:\;\; time--component--of}--\; g = (1-2 M
/ r)^{-1} = 0.
\end{array}
\label{eq:seven-19}\end{equation}
As a result, rules~(\ref{eq:seven-18}) contain conventional horizons and
singularities.

However, gravitational horizons and singularities are some of the most
significant  cases   of {\sl  interior}  gravitational  problems, thus
having nonlinear,  nonlocal and  nonlagrangian  effects indicated   in
Sect. 1   which are  outside  the representational  capability  of the
Riemannian  geometry. Rules~(\ref{eq:seven-18})  therefore  permit novel
studies on
gravitational horizons   and  singularities  with  a more    realistic
inclusion of    internal, velocity--dependent    and  nonlocal--integral
effects which will be conducted elsewhere.

It   should also be mentioned that    gravitational horizons are today
studied via  the use of {\sl  conventional} light cones, which implies
the assumption of  light {\sl  in vacuum}  $c_0$. But the  exterior of
gravitational horizons is made up of hyperdense chromospheres in which
the speed of electromagnetic waves  is a locally varying quantity. The
use of our isolight cone then permits more realistic calculations with
actual speeds of light, also contemplated in future works~[5e].

OIG   also permits the  resolution  of  the  {\sl  paradox of  quantum
mechanics  for gravitational  collapse},  identified  and  resolved in
ref.~[5d]. Consider a  conventional QM particle  in the interior of  a
star. As such,  the particle obeys conventional uncertainties. Suppose
now that the  star collapses all the  way to a singularity. Then,  the
particle considered must be located  at the singularity. But the  star
is  a {\sl classical}   object.  The location  of its  singularity can
therefore be classically determined  with the desired precision.   The
paradox of quantum mechanics here considered then follows, because the
QM particle should have uncertainties  at a point which is classically
determined.

It  seems evident that  no known resolution of  this paradox exists in
ordinary RQM. On the contrary, OIG  can indeed resolve the paradox. In
fact, the isoexpectation value of the commutation rules reads

\begin{equation}
<[\hat{r},\hat{}\hat{p}]> =
\frac{<|\times\hat{T}_{gr}\times(\hat{r}\times
\hat{T}_{gr}\times p -
\hat{p}\times\hat{T}_{gr}\times\hat{r})\times\hat{T}_
{gr}\times|>}{<|\times{T}_{gr}\times|>}
\label{eq:seven-20}\end{equation}
where the numerator is of the 4--th order in $\hat{T}_{gr}$ while the
denominator is of
the 1--st order. From rule~(\ref{eq:five-20}) it is then evident that

\begin{equation}
{\rm Lim}(\Delta r\times\Delta p)_{\hat{T}_{gr}\rightarrow 0} =
\frac{1}{2}{\rm Lim}<[\hat{r},\hat{}\hat{p}]>_{\hat{T}_{gr}\rightarrow
0}
\equiv 0,
\label{eq:seven-21}\end{equation}
namely, {\sl the isouncertainties acquire their classical deterministic
value
at the limit of gravitational collapse to a singularity}, which is the
main
result in of ref.~[5d] (other results in the same papers should be
reformulate
with the subsequent isodifferential calculus to avieve invariance, as
done in this study).

The above findings provide additional elements of plausibility
for OIG and indicate that the "completion" of RQM for the inclusion of
gravitation along the historical  E--P--R teaching~[14] is deeper that
just  the the achievement of an  explicit operator  realization of the
theory of "hidden variables"~, because  it implies rather  subtle
revisions of current studies on the relationship between classical and
operator theories  {\sl tacitly} restricted  to  the simplest possible
unit $\hbar=1$.

As  an example,   recall  that OIG  is  an  image  of   RQN under {\sl
nonunitary  transforms}  which, as such,    are known not  to preserve
eigenvalues  and  boundary values. It then  follows  that the image of
Bell's inequalities which is applicable to OIG is necessarily given by
a {\sl   nonunitary}  image of   the  conventional  forms~[14d].  This
evidently implies the  {\sl alteration of  the numerical value  of the
conventional  upper boundary of the inequalities   which, as such, can
become indeed compatible  with conventional classical counterparts for
the interior problem only} (see~[4h], App.~4.C for details, including
the construction  of the nonunitary  image of Pauli's  matrices). Note
the restriction, again, of the latter arguments  to the {\sl interior}
classical and operator problems. In fact, the {\sl exterior} classical
problem remains the conventional Hamiltonian--unitary one.

In  summary,  we  can  state that,  subject  to  further   studies and
independent scrutinies, OIG appears to resolve the problematic aspects
of QG, i.e.,  the lack of  invariant unit with consequential ambiguous
applicability to measurements, the lack of conservation of Hermiticity
in time with  consequential lack of physically acceptable observables,
the lack of invariance of the numerical predictions, physical laws and
special  functions   with consequential  lack  of  consistent physical
applications.

Moreover, QG is known to have  difficulties in achieving a formulation
comparable to ordinary RQM because of serious technical problems, e.g.,
in the construction of PCT symmetry. These difficulties are removed {\sl
ab initio}  in OIG precisely thanks  to  its {\sl isotopic} character,
that  is,  the inherent capability  of preserving   {\sl all} original
characteristics, as illustrated above.

Besides the transparent axiomatic advantages of OIG as compared to QG,
the plausibility of QIG is further established by {\sl preservation of
conventional quantum laws}, such as the conventional uncertainties for
the center--of--mass trajectories, Eq.s~(\ref{eq:seven-9}).

The latter property provide evident {\sl experimental} support for our
isominkowskian formulation  of  gravity,  this time,  at  the operator
level, because it indicates its  compatibility with existing,  quantum
mechanical experimental  data, a property which does  not appear to be
verified   by  QG. The  understanding   is  that contributions at  the
particle level due  to gravity are notoriously  very small as compared
to those due   to   other interactions,  as  established  by  isotopic
elements of the Schwarzschild's type~(\ref{eq:two-18}) where M is now
the mass of
the particle considered. These very  small contributions can therefore
be verified via suitable tests only after achieving the necessary
technology.

\vspace*{2cm}
\centerline{\bf \large Acknowledgments}

\vspace*{0.5cm}

The autor would like to  thank for invaluable comments the participants
to: the {\sl 7 Marcel Grossmann Meeting  on General Relativity} held at
Stanford university in July   1994; the {\sl International  Workshops}
held at the Istituto   per la Ricerca di  Base  in Molise,   Italy, on
August  1995 and  May 1996;  and the  {\sl  International Workshop  on
Physical Interpretation of Relativity  Theories} held at the  Imperial
College, London, on September 1996. Moreover,
the author would like to express his sincere appreciation
to the Editors of
{\it Rendiconti Circolo Matematico Palermo},
{\it Foundations of Physics} and {\it Mathematical Methods in Applied
Sciences},
for invaluable, penetrating and
constructive critical comments in the editorial processing of
the respective memoirs [3f,3g,7a] without which this paper could not
have seen
the light.

\end{document}